\documentclass[aps,prd,10pt,onecolumn,nofootinbib,showpacs,showkeys,notitlepage]{revtex4-1}
\usepackage{amsmath,graphicx,bm}

\renewcommand{\d}{\partial}
\newcommand{\dslash}{\mbox{$\d\hspace*{-0.5em}\raisebox{0.1em}[0.1em]{/}\hspace*{0.1em}$}}
\newcommand{\nn}{\nonumber\\}

\newcommand{\ph}{\varphi}

\newcommand{\exv}[1]{\left\langle{#1}\right\rangle}

\newcommand{\ep}{\varepsilon}

\newcommand{\T}{\textrm{T}}
\newcommand{\Tr}{\mathop{\textrm{Tr}}}

\def\pint4p{\int_p}

\newcommand\lsim{\mathrel{\rlap{\lower4pt\hbox{\hskip1pt$\sim$}} \raise1pt\hbox{$<$}}}                
\newcommand\gsim{\mathrel{\rlap{\lower4pt\hbox{\hskip1pt$\sim$}} \raise1pt\hbox{$>$}}}                

\begin{document}

\title{Strange mass dependence of the tricritical point \\in
  the $\bm{U(3)_L\times U(3)_R}$ chiral sigma model}

\author{A. Jakov\'ac}
\email{jakovac@fizika.phy.bme.hu}
\affiliation{Physics Institute, BME Technical University, 
H-1111 Budapest, Hungary}


\author{Zs. Sz{\'e}p}
\email{szepzs@achilles.elte.hu}
\affiliation{Statistical and Biological Physics Research Group
of the Hungarian Academy of Sciences, H-1117 Budapest, Hungary}     

\date{\today}

\begin{abstract}
  We study the strange quark mass dependence of the tricritical point
  of the $U(3)_L\times U(3)_R$ linear sigma model in the chiral limit.
  Assuming that the tricritical point is at a large strange mass
  value, the strange sector as well as the $\eta-a_0$ sector decouples
  from the light degrees of freedom which determines the
  thermodynamics.  By tracing this decoupling we arrive from the
  original $U(3)_L\times U(3)_R$ symmetric model, going through the
  $U(2)_L\times U(2)_R$ symmetric one, at the $SU(2)_L\times SU(2)_R$
  linear sigma model.  One-loop level beta functions for the running
  of the parameters in each of these models and tree-level matching of
  the coupling of these models performed at intermediate scales are
  used to determine the influence of the heavy sector on the
  parameters of the $SU(2)_L\times SU(2)_R$ linear sigma model.  By
  investigating the thermodynamics of this latter model we identified
  the tricritical surface of the $U(3)_L\times U(3)_R$ linear sigma
  model in the chiral limit. To apply the results for QCD we used
  different scenarios for the $m_s$ and $\mu_q$ dependence of the
  effective model parameters, then the $\mu_q^\textnormal{TCP}(m_s)$
  function can be determined. Depending on the details, a curve
  bending upwards or downwards near $\mu_q=0$ can be obtained, while
  with explicit chemical potential dependence of the parameters the
  direction of the curve can change with $m_s$, too.
\end{abstract}

\maketitle

\section{Introduction}

The phase diagram of QCD is a much-studied phenomenon, but still its
characteristics, at finite baryon chemical potential in particular,
are far from being settled \cite{deForcrand:2010ys, Fodor:2009ax,
  Fukushima:2010bq}. While at zero chemical potential all the Monte
Carlo (MC) and effective model studies tend to support a common
picture, at nonzero chemical potential the MC results are
inconclusive.

At zero chemical potential the widely accepted phase diagram in the
$m_{ud}$-$m_s$ plane exhibits both at small and at large quark masses
regions of first order phase transition each bounded by a line of
second order critical endpoints (CEP). In between these regions, the
transition is of analytic crossover type. If we introduce a nonzero
quark baryon chemical potential $\mu_q$, the second order CEP lines
extend to a critical surface. If the critical surface lying closer to
the origin of the mass-plane bends upwards, that is to larger quark
masses, then there is a possibility that a crossover transition
becomes at larger chemical potential a second, and subsequently a
first order phase transition. Direct MC simulations
\cite{Fodor:2004nz, Csikor:2004ik} and also the estimates in
\cite{Ejiri:2003dc,Li:2010dy} give a finite $\mu_q^\text{CEP}$ value,
and similar conclusions can be drawn from other lattice techniques,
too \cite{Ejiri:2006ft, Gavai:2008zr,Allton:2005gk}.  However, the
second order surface seems to bend downwards, to smaller quark masses,
according to a lattice study of the curvature performed in
\cite{deForcrand:2008zi, deForcrand:2006pv, deForcrand:2007rq} using
imaginary chemical potential.  Although all these lattice results do
not necessarily contradict each other, they could imply a scenario in
which the critical surface has a non-trivial shape. Some numerical
evidence for such a possibility is given in \cite{deForcrand:2010he}.

Beyond direct simulations one can approach the study of the phase
structure of QCD through effective theories, see {\it e.g.}
\cite{Casalbuoni:2006rs,Fukushima:2010bq} for reviews.  At finite
chemical potential there are several results on the chiral phase
transition for two flavors in Nambu--Jona-Lasinio (NJL) models, as
well as in linear sigma models \cite{Berges:1998rc, Scavenius:2000qd,
  Jakovac:2003ar, Bowman:2008kc}.  Using the $SU(2)_L\times SU(2)_R$
linear sigma model an interesting phase structure with two CEPs in the
$\mu_q-T$ plane was reported for low values of the pion mass in
\cite{Bowman:2008kc}.  Considering three flavors, one can study in
these models the properties of the chiral critical surface
\cite{Kovacs:2006ym, Kovacs:2007sy,Schaefer:2008hk, Fukushima:2008wg,
  Fukushima:2008is, Chen:2009gv}. In \cite{Kovacs:2006ym,
  Kovacs:2007sy,Schaefer:2008hk} the authors used the $U(3)_L\times
U(3)_R$ chiral sigma model near the physical point, and found that in
the available parameter space the critical surface bends upwards,
supporting the direct MC result. In \cite{Fukushima:2008wg,
  Fukushima:2008is, Chen:2009gv}, in the framework of the extended NJL
model the authors found a down-bending surface for small chemical
potential which eventually turns back at higher values of $\mu_q.$
This behavior would conciliate the two MC scenarios within a single
critical surface if the turning happens at positive values of
$m_{ud},$ $m_s$ and finite values of $\mu_q$. Other possible behaviors
of the critical surface were discussed in
\cite{Gupta:2007dx,Gupta:2008ac} using the Gibbs' phase rule for phase
coexistence.

In the chiral limit ($m_{ud}=0$), there are two well-known limits, the
$\mu_q=0$ and the infinite strange quark mass ($m_s=\infty$)
limits. At $\mu_q=0$ we have a first order phase transition region for
small $m_s$, a second order transition region at large $m_s$ and a
tricritical point (TCP) separating them, with a characteristic value
$m_s^\text{TCP}.$ At $m_s=\infty$ and small chemical potential we have
second order, for large $\mu_q$ a first order phase transition, with
again a TCP in between. The line of TCPs is the intercept of the
critical surface and the $m_s-\mu_q$ plane. According to the above
scenarios the two TCPs at the $\mu_q=0$ and $m_s=\infty$ can be
connected by a single line with definite curvature, a back-bending
line, or they may belong to two distinct TCP lines. In the latter case
there must be two disjunct critical surfaces in the $m_{ud}-m_s-\mu_q$
space, of which these two TCP lines are just the endpoints in the
$m_s-\mu_q$ plane \cite{Gupta:2007dx,Gupta:2008ac}.

In this paper we attempt to describe the behavior of the tricritical
line in the chiral limit of the $U(3)_L \times U(3)_R$ sigma model
\cite{Lenaghan:2000ey, Herpay:2006vc, Kovacs:2006ym}, assuming in
addition that the mass of the constituent strange quark and the
anomaly scale are much larger than the critical temperature. The study
of the chiral limit has some advantages, since one can work with much
less parameters than in a generic situation. A disadvantage though, is
that in this case there are no direct measurements which could connect
the effective model parameters with the QCD (although, strictly
speaking, there are no such measurements anywhere, apart from the
physical point, especially not for a possible chemical potential
dependence of the parameters). The goal of this study is to explore
the parameter dependence of the TCP line.

The assumption that the constituent strange mass is much larger than
$T_c$ is based on the observation that even at the physical point of
the mass plane the constituent strange quark mass is $\sim450$~MeV,
while $T_c\sim 160$~MeV, and the critical line in the $m_{ud}-m_s$
plane behaves as $m_s^\text{TCP}-m_s\sim m_{ud}^{2/5},$ as one
approaches the $m_s$ axis, which predicts for $m_s^\text{TCP}$ and for
the mass of the strange constituent quark a higher value than the
corresponding mass at the physical point. Actually, in the lattice
study of \cite{Philipsen:2005mj} it was estimated that
$m_s^\text{TCP}\simeq 2.8T_c,$ while in \cite{Herpay:2006vc} using the
$U(3)_L \times U(3)_R$ sigma model $m_s^\text{TCP}$ was estimated to
be one order of magnitude bigger than the value of $m_s$ at the
physical point.  A similar observation can be made for the anomaly
scale, which is connected to the mass of the $\eta'$ meson.  In this
physical situation we have a multi-scale system, where a simple
one-loop analysis would lose the contribution of the heavy
sector. Instead, we have to work with decoupling theory \cite{Collins}
which results in a hierarchy of effective models describing the
physics at lower and lower scales: first the $m_s$ strange quark
sector decouples with the corresponding bosonic degrees of freedom,
and we obtain an effective $U(2)_L\times U(2)_R$ symmetric
theory. Then the $\eta'$ sector (which has semi-large masses at
$m_s\to\infty$ because of the anomaly) decouples, and we are left with
the $SU(2)_L\times SU(2)_R$ chiral sigma model consisting of pion and
sigma mesons, as well as the $u$ and $d$ constituent quarks. Here we
can use the results of \cite{Jakovac:2003ar} to determine the position
of TCP for a given parameters set.  The effect of the strange sector
on the position of the TCP is only through the modified parameters of
the $SU(2)_L\times SU(2)_R$ chiral sigma model. To determine the
parameters of the different effective models involved in the analysis
we use the one-loop $\beta$-function governed running of these
parameters and matching of the corresponding $n$-point functions in
the common validity range of the models. Some extra assumptions on the
original parameters of the $U(3)_L \times U(3)_R$ sigma model are
unavoidable.

The setup of the paper is as follows. First, we discuss the model in
Sec.~\ref{sec:model}. Then, we overview the generic ideas how the
decoupling works in Sec.~\ref{sec:dec}. Next, we perform the
decoupling in the $U(3)_L\times U(3)_R$ model in Sec.~\ref{sec:U3}. We
study the thermodynamics in the resulting effective 
$SU(2)_L\times SU(2)_R$ linear sigma model in Sec.~\ref{sec:thermo}. 
We close with conclusions in Sec.~\ref{sec:conc}.

\section{The model}
\label{sec:model}

We first construct the starting model \cite{Lenaghan:2000ey,
  Kovacs:2006ym}, which is the $U(3)_L\times U(3)_R$ symmetric linear
sigma model defined by the Lagrangian:
\begin{equation}
  \label{NLag}
  {\cal L}_{U(3)} = \bar \psi \bigl[i\dslash - 2g
  T^a(\sigma_a+i\gamma_5\pi_a)\bigr] \psi + \Tr(\d_\mu \Phi^\dagger
  \d_\mu \Phi) - U(\Phi),
\end{equation}
where $T^a$ are the $U(3)$ generators, $T_a=\lambda_a/2$ for
$a=1,\dots 7$ and $T^0\equiv T^x,\, T^8\equiv T^y$
\cite{Herpay:2006vc}, with
\begin{equation}
  T^{x}=\frac12 \left(\begin{array}[c]{rrr}
      1\;&0\;&0\cr
      0\;&1\;&0\cr
      0\;&0\;&0\cr
    \end{array}\right)
  \quad
  T^{y}=\frac1{\sqrt{2}} \left(
    \begin{array}[c]{rrr}
      0\;&0\;&0\cr
      0\;&0\;&0\cr
      0\;&0\;&1\cr
    \end{array}\right).
\end{equation}
The meson matrix $\Phi=\bm \sigma+ i\bm \pi =T^a(\sigma_a+i\pi_a)$ 
is given in terms of the physical degrees of freedom as:
\begin{eqnarray}
  &&{\bm\sigma} = \frac1{\sqrt{2}} \left(
    \begin{array}[c]{ccc}
      \frac1{\sqrt{2}}(\sigma_x+a_0^0) \quad& a_0^+\quad& \kappa^+\cr
      a_0^-\quad&\frac1{\sqrt{2}}(\sigma_x-a_0^0)\quad& \kappa^0\cr 
      \kappa^-\quad& \bar\kappa^0\quad& \sigma_y\cr
    \end{array}\right),\qquad
  {\bm\pi} = \frac1{\sqrt{2}} \left(
    \begin{array}[c]{ccc}
      \frac1{\sqrt{2}}(\eta_x+\pi^0) \quad& \pi^+\quad& K^+\cr
      \pi^-\quad&\frac1{\sqrt{2}}(\eta_x-\pi^0)\quad& K^0\cr 
      K^-\quad& \bar K^0\quad& \eta_y\cr
    \end{array}\right).
\end{eqnarray}
Finally, the potential for $\Phi$ reads
\begin{equation}
  \label{SUNSUNLag}
  U(\Phi) = M^2 \Phi^\dagger \Phi + \lambda_1 \big[\Tr(\Phi^\dagger \Phi)\big]^2 + \lambda_2 \Tr\big[(\Phi^\dagger \Phi)^2\big] -\sqrt{2} C (\det \Phi+\det \Phi^\dagger) + \Tr \big[H(\Phi+\Phi^\dagger)\big].
\end{equation}
Here, $C$ governs the $U(1)$ anomaly which breaks the symmetry to
$SU(3)_L\times SU(3)_R.$ The last $H$-dependent term explicitly breaks
also this symmetry. As a consequence we assume that vacuum expectation
values develop for those scalar fields belonging to the center
elements of the symmetry group. These are taken into account through
the shifts: $\sigma_x\to\sigma_x+x$ and $\sigma_y\to\sigma_y+y$. In
this paper we are interested in the regime where the vacuum structure
of the theory contains a heavy $s$-quark sector and a light $ud$
sector, with a corresponding meson sector. For the constituent quark sector 
this mass hierarchy can be fulfilled with $y\gg x$, since at tree-level the
$s$-quark mass is $m_s=\sqrt{2} gy$, while $m_{ud}=gx$. We assume that
the mesons containing $s$-quark have a mass of order $m_s,$ while the
rest have a mass of order $m_{ud}.$ No splitting is assumed between
$u$ and $d$ quark masses.

After having made all these assumptions, the mass spectrum, and
correspondingly the physics splits into a heavy and a light part.  The
light sector influences only very little the symmetry breaking pattern
of the heavy sector, therefore we may set $x=0$ when dealing with the
determination of the heavy masses. Then we find the following mass
relations:
\begin{eqnarray}
  \label{masses1}
  && m_x^2 = M^2+\lambda_1 y^2-C y,\quad
  m_a^2 = M^2+\lambda_1 y^2+C y,\quad
  m_K^2 = M^2+(\lambda_1+\lambda_2) y^2,\nn&&
  m_y^2 = M^2+3(\lambda_1+\lambda_2) y^2,\quad
  m_s^2 = 2g^2 y^2,
\end{eqnarray}
where $m_x^2$ is the mass for $\pi_i$ and $\sigma_x$, $m_a^2$ is the
mass of $\eta_x$ and $a^0_i$, $m_K^2$ applies for $\eta_y,\,K$ and
$\kappa$ modes, $m_y^2$ is the mass of $\sigma_y$ and finally, $m_s$ is
the strange quark mass. In case of a large strange quark mass, {\it i.e.} large
$y$ values, the lightness of the pion-sigma sector requires that
$m_x^2\ll m_s^2$. This can be achieved only with fine-tuning $m^2$ --
after all this is the manifestation of the hierarchy problem, where
the high energy sector influences, through radiative corrections and
spontaneous symmetry breaking, the light sector. To circumvent the
problem, we parametrize everything with the light $x$-mass:
\begin{equation}
  \label{masses2}
  m_a^2 = m_x^2+2C y,\quad m_K^2 = m_x^2+ \lambda_2 y^2,\quad m_y^2 =
  m_x^2+(2\lambda_1+3\lambda_2) y^2.
\end{equation}
We see that there is a third mass scale, associated to $C y$. Since
$C$ has the dimension of a mass, we can relate its value to $y$ or
$m_x$, and it is a matter of the low-lying dynamics determining the
details of the effective model to know which is the ``true'' ratio
between them. In this work we try to play around the possible values
of $C$ to see its effect on the thermodynamics.

So, after all, we have three possibly very different mass scales,
$m_s,\, m_a$ and $m_x$. The thermodynamics of the system, which is
related to the spontaneous breaking in the non-strange sector, must take
place at the light scales, that is at $T\sim m_x$. To treat a physical
system with vastly different mass scales is possible only using the
fact that for the light physics the heavy degrees of freedom decouple,
and their presence can be identified through the values of the
parameters of the Lagrangian containing the light degrees of
freedom. How this decoupling works in detail, is summarized in the
next section.

\section{Decoupling of mass scales}
\label{sec:dec}

Here we review the generic principles of decoupling described in
details in \cite{Collins}. Let us start with a theory with Lagrangian
${\cal L}$, coupling constant set $\{g_i\}_{i=1\dots u}$ (this
includes also the masses) and field contents generically denoted with
$\Phi$ and $\varphi$. Let us assume that $\Phi$ is much heavier than
$\ph$, their masses are denoted by $M$ and $m$,
respectively\footnote{For the sake of simplicity we only consider one
  heavy and one light degree of freedom}. Then we want to establish an
effective theory based exclusively on the light degrees of
freedom. Let us denote the Lagrangian of this effective theory by
$\hat{\cal L}$, its couplings are $\{\hat g_i\}_{i=1\dots\hat u}$
($\hat u\le u$) and the field content is $\hat\ph$ with mass $\hat m
\sim m$. We emphasize that, although $\hat\ph$ corresponds to the
light degrees of freedom $\ph$ of the complete theory, they can differ
by wave function renormalization. In general the wave function
renormalization and the renormalized couplings of the effective theory
depend on the coupling constants of the original theory:
\begin{equation}
  \label{rel}
  \hat\ph = z^{1/2}(g) \ph,\qquad \hat g_i= f_i(g).
\end{equation}
If we assume that the couplings of the effective model at tree level
are linear combinations of the original couplings, then we can write
\begin{equation}
  \hat g_i = \sum\limits_{j=1}^u {\cal G}_{ij}g_j + \Delta
  g_i(g),\qquad i\in[i\dots \hat u],
\end{equation}
where $\Delta g_i(g)$ denotes the loop corrections which depend on all
of the original couplings.

In order to determine these relations we use matching: we take $\hat
u+1$ correlators of the light fields at some external momentum, and
require that their physical value be the same in the original and in
the effective model. If we denote the $n$-point functions of the
original and effective model by
\begin{equation}
  G^{n}(x_1,\dots x_n) = \exv{\T \ph(x_1)\dots \ph(x_n)},\qquad
  \hat G^{n}(x_1,\dots x_n) = \exv{\T \hat\ph(x_1)\dots \hat\ph(x_n)},
\end{equation}
then we require in Fourier space 
\begin{equation}
  \label{match1}
  \hat G^{n}(k_1,\dots k_n) = z^{n/2} G^{n}(k_1,\dots k_n)
\end{equation}
for fixed $k_1,\dots k_n$. The right hand side contains, as radiative
corrections, the effect of heavy modes.

In perturbation theory the $n$-point functions depend also on the
renormalization scales $\mu$ and $\hat\mu$, respectively. In general,
we expect to obtain all type of logarithmic corrections $\ln^a\mu/E$,
where $E$ is any energy scale which shows up in the given $n$-point
function. In loop integrals we typically find $E^2 \sim
\max(k_i^2,\mbox{mass}^2)$, where the mass can be $m,\, M$ on the
right, or $\hat m$ on the left hand side. In order not to have
multiple mass scales in the $n$-point functions (which would lead to
large logarithms) we shall choose $|k_i|\sim M$, then all scales are
of the order of $M$. This means that for the best convergence of the
perturbative series we shall also choose $\mu,\, \hat\mu\sim M$.
In consequence the relation \eqref{rel} is in fact
established at scale $M$:
\begin{equation}
  \label{rel1}
  \hat g_i(M) = \sum\limits_{j=1}^u {\cal G}_{ij} g_j(M) + \Delta g_i(g(M)).
\end{equation}
In the original theory the couplings may be defined on a different
energy scale, $M_0$. Then, we have to run the couplings according to
${\cal L}$ in order to find $g_i(M)$ which is needed above. On the
other hand, \eqref{rel1} \emph{defines} $\hat g_i$ on the scale
$M$. If we need them on a lower energy scale $M'$, then we have to run
them according to $\hat{\cal L}$, the effective model Lagrangian.

\begin{figure}[htbp]
  \centering
  \includegraphics[height=4cm]{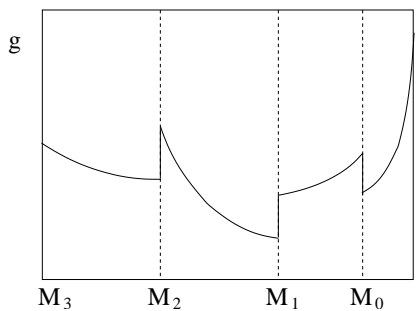}
  \caption{Schematic plot of the running of a coupling constant. The
    running between $M_{(a)}$ and $M_{(a+1)}$ is governed by ${\cal
      L}_{(a+1)}$.}
  \label{fig:schemplot}
\end{figure}

If there are multiple scales to decouple, we have a series of
effective models ${\cal L}^{(a)}$ with coupling constants
$\{g^{(a)}_i\}_{i=1\dots u_a}$ where the heaviest field has a mass
$M_{(a)}$. Then the matching conditions described above lead to the
series of equations
\begin{equation}
  g_i^{(a+1)}(M_{(a)}) = \sum\limits_{j=1}^{u_a} {\cal G}_{ij}
  g_i^{(a)}(M_{(a)}) + \Delta g_i^{(a)}(g(M_{(a)})).
\end{equation}
This defines the new couplings $g_i^{(a+1)}$ at the mass scale
$M_{(a)}$ which, from the point of view of the $a+1$th effective
model, is a high energy scale. Then the running of the couplings
between $M_{(a)}\to M_{(a+1)}$ is governed by the Lagrangian ${\cal
  L}^{(a+1)}$. This process leads to the schematic running depicted
on Fig. \ref{fig:schemplot}.

In the most simple version we use tree level matching, which means
that we neglect the $\Delta g_i$ terms. We use this approximations
throughout this work. Moreover, we use the lowest order (one-loop)
beta functions for running.

We start with the $U(3)_L\times U(3)_R$ symmetric linear sigma model
which is parametrized at the heavy, strange mass scale in order to
avoid the need for running. After the decoupling of the strange sector
we are left with a $U(2)_L\times U(2)_R$ model, starting at $m_s$
scale. The beta-functions of this model determines the running of the
coupling down to $m_a$, when we switch to $SU(2)_L\times SU(2)_R$
model containing only the $\sigma$ and $\pi$ mesons and the $u$ and
$d$ constituent quarks. We perform in this model the running of their
couplings from $m_a$ down to $T$ scale, where the finite temperature study
is performed.

A separate treatment is needed when we treat the light masses of the
system. At one hand, due to the RG running, they acquire logarithmic
dependence on the heavy scale. On the other hand, because of the
hierarchy problem discussed in the previous section, the light mass
squared $m^2$ have to be fine tuned by $M^2$ in order to keep the
light sector truely light. Therefore in bosonic models the logarithmic
corrections from the RG running are subleading, and so they should be
neglected.

\section{Decoupling of the heavy sector}
\label{sec:U3}

We use \eqref{NLag} with a background $y$ taken into account via the
shift $\sigma_y\to\sigma_y+y$. The first step is to parametrize the
heavy sector by determining the parameters of the model at scale
$m_s$.  Then, as a next step, we perform the decoupling of the heavy
(strange) sector.

\subsection{Parametrization of the heavy sector
\label{ss:heavy_para}}

We should fix the parameters of the heavy sector by measurements, but
there exist no direct mass measurements in this regime. Therefore we
are forced to make some assumptions. Since we are in a large quark
mass regime, we can not use the results of the chiral perturbation
theory. Instead, the heavy constituent quark model approach, where the
mass of the heavy particles are simply the sum of the constituent
quark masses, seems to be more adequate.  In case of the light-heavy
mesons this works nicely, since we can require
\begin{equation}
  m_K = m_s \qquad\Rightarrow\qquad \lambda_2 = 2g^2.
\end{equation}
In the doubly heavy sector there is a mismatch between the scalar
($\sigma_y$) and pseudoscalar ($\eta_y$) meson masses, although the
constituent quark model would give the same mass for both. The reason
in this model is that there is a symmetry breaking effect in addition
to the constituent quark masses. To treat this situation we introduce
a free parameter $\bar{\cal A}$, and require that some average of the
two mass squared is $(2m_s)^2$:
\begin{equation}
  \bar{\cal A} m_{y}^2 + (1-\bar{\cal A}) m_K^2 = (2m_s)^2,
\end{equation}
which results in a relation between the two quartic coupling constants
$\lambda_1$ and $\lambda_2$
\begin{equation}
 \lambda_1 = \left(\frac3{2\bar{\cal A}}-1\right) \lambda_2.
\end{equation}
If $\bar{\cal A}=1$, {\it i.e.} when $m_{y} = 2m_s$, $\lambda_1=g^2$
is the smallest value for $\lambda_1$. Another plausible choice is 
$\bar{\cal A}=1/2$, then $\lambda_1 = 2\lambda_2 = 4g^2$. Therefore, 
introducing ${\cal A}=-2+3/\bar{\cal A}$ we may set
\begin{equation}
  \lambda_1 = {\cal A} g^2,\qquad\mathrm{where}\quad {\cal A}\in[1,\infty].
\end{equation}
The parameter range ${\cal A}\in[1,4]$ will be used later in the analysis.

For fixing the other parameters we will use the infrared (IR) sector:
we will determine $m_{\sigma}$ at its own scale, this will give the
mass unit in the study. For fixing the $m_s$ value at $\mu=m_s$ scale
we also use an IR observable: this will be that $m_s^{(0)}$ value
where we have a TCP at \emph{zero chemical potential} (but, of course,
at finite temperature). Through the decoupling equations this value
will determine the original, zero temperature strange quark mass.

\subsection{Tree level decoupling}

The next step is to eliminate the strange sector, and determine the
parameters of the resulting effective model. The effective model will
contain the following degrees of freedom: the upper two ($u,\,d$)
components in $\psi$, denoted by $\psi_2,$ and the upper left
$2\times2$ submatrix in $\Phi,$ denoted by $\Phi_2$ and containing the
$\sigma\equiv\sigma_x,\, a_0,\, \eta_x,$ and $\pi$ fields:
\begin{equation}
  \psi=\left(\begin{array}[c]{c}\psi_2\cr s\cr \end{array}\right),\qquad
  \Phi =   \left(\begin{array}[c]{cc} \Phi_2 \;& \displaystyle
      \frac1{\sqrt{2}} {\bm K}_+\cr \displaystyle \frac1{\sqrt{2}}{\bm
        K}_-^\dagger \quad& \displaystyle \frac1{\sqrt{2}}(\sigma_y+i\eta_y)
      \cr\end{array}\right),\qquad {\bm K}_\pm = {\bm\kappa}\pm i{\bm K}
  =\left( \begin{array}[c]{c} \kappa^+\pm iK^+ \cr \kappa^0\pm
      iK^0 \end{array}\right). 
\end{equation}
We split the Lagrangian \eqref{NLag}-\eqref{SUNSUNLag} according to
this characterization of degrees of freedom. The shifted Lagrangian
reads
\begin{equation}
  \label{U3U3shift}
  {\cal L}_{U(3)} = {\cal L}_{U(2)} + {\cal L}_\text{heavy} +\mbox{const.}\,.
\end{equation}
Here,  ${\cal L}_{heavy}$ contains the heavy part, the `$\mbox{const.}$' term 
refers to the $y$-dependent part of the potential, while ${\cal L}_{U(2)}$ 
contains the terms which consist of the light fields:
\begin{eqnarray}
  \label{L2}
  {\cal L}_{U(2)} =&& \bar\psi_2 \left[ i\dslash -
    g\tau_i(\sigma_{2i}+i\gamma_5 \pi_{2i}) \right] \psi_2 + 
  \Tr \big(\d_\mu \Phi_2^\dagger \d_\mu \Phi_2  \big)
  - (M^2+\lambda_1 y^2) \Tr\big[\Phi_2^\dagger \Phi_2\big]\nn&& 
  -\lambda_1 \big[\Tr (\Phi_2^\dagger \Phi_2)\big]^2
  -\lambda_2\Tr\big[(\Phi_2^\dagger \Phi_2)^2\big] 
  + C y \left[\det \Phi_2 + \det \Phi_2^\dagger\right].
\end{eqnarray}

Tree level matching means that we simply neglect the heavy part, and
go on with the Lagrangian described above. We can argue for this
simple choice as follows. The parametrization of the heavy sector of
the model could be done only very heuristically, in which case only
the leading order effects could be taken into account. Therefore, it
would be inconsistent to work with a detailed decoupling scheme,
determining $\Delta g\sim {\cal O}(g^2)$ or ${\cal O}(g^3)$
corrections to the tree level values which are not known
precisely. For this reason the leading order approach is the most
consistent here. As a result of this approximation we shall trust only
the most robust consequences of this study.

\subsection{Running in the $U(2)_L\times U(2)_R$ linear sigma model}

In component fields \eqref{L2} can be written as
\begin{eqnarray}
  \label{L2c}
  {\cal L}_{U(2)} =&& \bar \psi_2 \bigl[i\dslash - g(\ph_5-i\gamma_5 a_5)
  \bigr]\psi_2 + \frac12 (\d_\mu\ph)^2 -\frac{m_\varphi^2}2\ph^2 + \frac12
  (\d_\mu a)^2 -\frac{m_a^2}2a^2 \nn&& -\frac\lambda4(\ph^2+a^2)^2
  -\frac{\lambda_2}2(\ph^2a^2-(\ph a)^2),
\end{eqnarray}
where we introduced the following notations:
$\varphi_5=\sigma_0+\tau_i\sigma_i,$ $a_5=-\pi_0-\tau_i\pi_i,$ with
$\tau_i$ being the Pauli matrices, $\lambda=\lambda_1+\lambda_2/2$,
$\ph=(\sigma,\pi_i)$, and $a=(-\eta,a_i).$ The squared masses are
$m^2_{\varphi/a}= m^2\mp c,$ where $m^2=M^2+\lambda_1 y^2$ and $c=C y$
(cf.~\eqref{masses1}).

The running of the couplings are determined by the beta-functions
given in Appendix~\ref{sec:U2}, under equation \eqref{U2beta}.  By
solving (\ref{Eq:U2_g_run}) the running of $g$ can be obtained
explicitly:
\begin{equation}
  \label{U2run}
  g^2(\mu) = \frac{g^2(\mu_0)} {  1-\displaystyle
    \frac{5g^2(\mu_0)}{12\pi^2} \ln\frac{\mu^2}{\mu_0^2} } =
  \frac{12\pi^2} {\displaystyle 5\ln \frac{\bar\Lambda_0^2}{\mu^2}},
\end{equation}
where $\bar\Lambda_0^2 = \mu_0^2 \exp{\left[12\pi^2/(5g^2(\mu_0))\right]}$.
This has an UV Landau pole, while it goes to zero when $\mu\to 0$.

\begin{figure}[!t]
  \centering
  \includegraphics[keepaspectratio,width=0.5\textwidth,angle=0]{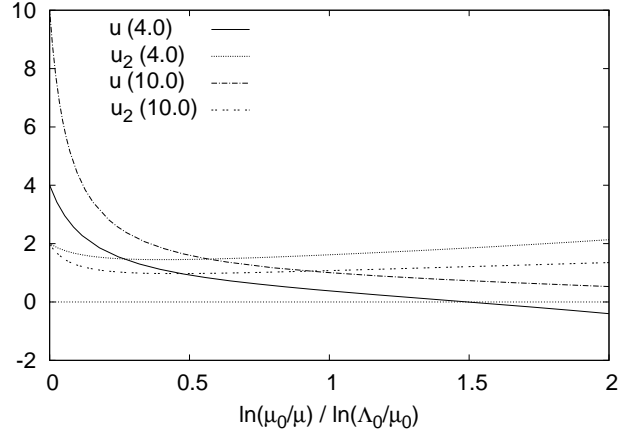}
  \caption{Running of the ratios $u=2\lambda/g^2$ and
    $u_2=\lambda_2/g^2$ starting from different initial conditions.}
  \label{fig:running}
\end{figure}

For the other two equations, (\ref{Eq:lambda_run}) and (\ref{Eq:lambda2_run}), 
we introduce the ratios:
\begin{equation}
  \label{ratios}
  u =\frac{2\lambda}{g^2},\qquad u_2=\frac{\lambda_2}{g^2},
\end{equation}
and a new function $X(\mu)$ monotonous in $\mu$ which satisfies
\begin{equation}
  \label{Xmu}
  \frac1{g^2}\,\frac{dX}{d\ln\mu}=\frac1{4\pi^2}.
\end{equation}
Using (\ref{U2run}), this equation has the solution
\begin{equation}
  X(\mu) = \frac3{10} \ln\left[\frac{\ln\bar(\Lambda_0/\mu_0)}
    {\ln(\bar\Lambda_0/\mu)}\right],\quad \ln\frac{\mu_0}\mu =
  \ln\frac{\bar\Lambda_0}{\mu_0} \left(e^{-\frac{10}3 X}-1\right) =
  \ln\frac{\bar\Lambda_0}\mu \left(1-e^{\frac{10}3X}\right).
\end{equation}
Here we have chosen the condition $X(\mu_0)=0$, where $\mu_0$ is chosen to 
be the strange quark mass $m_s$. Then we find:
\begin{eqnarray}
  \label{U2urun}
  &&\frac{\d u}{\d X} = 4u^2 + 3uu_2 +3u_2 -4-\frac{14}3u,\nn&&
  \frac{\d u_2}{\d X} =3u u_2 + u_2^2- 4-\frac{14}3u_2.
\end{eqnarray}
Conform subsection \ref{ss:heavy_para}, 
the phenomenologically motivated initial conditions are $u_2=2$ and
$u=2({\cal A}+1) = 4\dots 10,$ at the scale of the $s$-quark. 
The solution of (\ref{U2urun}) is depicted in Fig. \ref{fig:running}.
As this plot also demonstrates, for a wide range of $\mu$, $u_2$ stays
in the interval $[1,2]$, while $u$ decreases continuously as we
lower the scale $\mu$ from $\mu_0=m_s$ down to $\mu<m_s$. Sooner or
later (depending the initial conditions) $u$ crosses zero which
signals the instability of the theory. Probably it means that in order
to maintain stability higher order corrections are needed.

\subsection{Running in the $SU(2)_L\times SU(2)_R$ linear sigma model}

If we are well below the $m_a$ scale, then we can use the model
containing the $\sigma-\pi$ sector of \eqref{L2c} and the $u,d$
constituent quarks. This model is the $SU(2)_L\times SU(2)_R$ linear
sigma model defined by
\begin{equation}
  \label{LO4}
  {\cal L}_{SU(2)} =\bar \psi_2 \bigl[i\dslash - g(\sigma +
  i\gamma_5\tau_i\pi_i)\bigr]\psi_2 + \frac12 (\d_\mu\ph)^2
  -\frac{m_\varphi^2}2\ph^2 - \frac\lambda4 \ph^4,
\end{equation}
where $ \psi_2$ and $m_\varphi^2$ were defined in the previous two subsections.
The parameters of the model are defined at scale $m_a$, so we have to
apply renormalization group running to find the values of the coupling
at the phase transition temperature $T\sim m_\sigma$.

The RG equation are determined in Appendix~\ref{sec:O4},
eq. \eqref{Eq:beta_fn_S}.  The running of $g$ can be solved:
\begin{equation}
  \label{O4run}
  g^2(\mu) = \frac{g^2(\mu_0)} {  1+\displaystyle
    \frac{5g^2(\mu_0)}{24\pi^2} \ln\frac{\mu^2}{\mu_0^2} } =
  \frac{24\pi^2} {5\displaystyle \ln \frac{\mu^2}{\Lambda_0^2}},
\end{equation}
where $\Lambda_0^2 = \mu_0^2 \exp[-24\pi^2/(5g^2(\mu_0))].$ This has
an IR Landau pole, while it goes to zero when $\mu\to\infty$. The
definition of $\Lambda_0$ is RG invariant. Comparing (\ref{O4run})
with \eqref{U2run}, and taking into account that the change of scaling
is at $\mu_0=m_a$ we find
\begin{equation}
  \label{scalerel}
  \bar\Lambda^2_0 = \frac{m_a^3}{\Lambda_0}.
\end{equation}

\begin{figure}[!t]
  \centering
  \includegraphics[keepaspectratio,width=0.5\textwidth,angle=0]{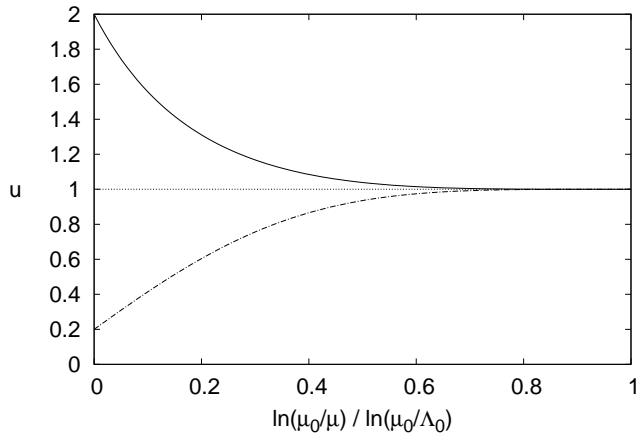}
  \caption{Running in the $SU(2)_L\times SU(2)_R$ linear sigma model 
    showing the presence of a fix point at $u=1$.}
  \label{fig:O4run}
\end{figure}

For the running of $\lambda$ we again introduce again $X(\mu)$ defined
in \eqref{Xmu} where now $g$ is the coupling of the $SU(2)_L\times
SU(2)_R$ linear sigma model.  Using (\ref{O4run}) we obtain
\begin{equation}
  X(\mu)=\frac35
  \ln\left[\frac{\ln(\mu/\Lambda_0)}{\ln(\mu_0/\Lambda_0)} \right],\quad
  \ln\frac{\mu_0}\mu = \ln\frac{\mu_0}{\Lambda_0} \left(1-e^{\frac53X}\right).
\end{equation}
We introduce the same ratio as in \eqref{ratios}: $2\lambda=ug^2$, and find
\begin{equation}
  \frac{du}{dX} = 3u^2+u-4.
\end{equation}
The solution of this equation reads
\begin{equation}
  \label{O4urun}
  \frac{u-1}{3u+4} =  \frac{u_0-1}{3u_0+4}
  \left(1-\frac{\ln(\mu_0/\mu)} {\ln(\mu_0/\Lambda_0)}\right)^{21/5}.
\end{equation}
This equation has a fixed point at $u=1$, as it is shown in
Fig. \ref{fig:O4run}

In the $SU(2)_L\times SU(2)_R$ linear sigma model we can follow the
running of the light mass, using (\ref{Eq:beta_fn_S}). The same
running is true for the tree-level sigma mass
$m_\sigma^2\equiv-2m^2_\varphi$. Then we find
\begin{equation}
  m_\sigma^2(\mu) = m_\sigma^2(\mu_0) \exp\left[{\frac1{12\pi^2}
  \int_0^{\ln\frac\mu{\mu_0}} d\left(\ln(\mu'/\mu_0)\right) 
  g^2(\mu')\left(\frac{9}{2}u(\mu')-1\right)}\right].
\end{equation}
With the help of \eqref{O4run} and \eqref{O4urun} we obtain
\begin{equation}
  \label{mxrun}
   m_\sigma^2(\mu) = m_\sigma^2(\mu_0) \exp\left[\frac{7g^2(\mu_0)}{24\pi^2}\!\!
     \int_0^{\ln\frac\mu{\mu_0}}\!\!\frac{ds}{1+ b g^2(\mu_0)s}
     \, \frac{1+6 Y_0 z(s)}{1-3 Y_0 z(s)}\right],
\end{equation}   
where 
$\displaystyle z(s) = \left(1+\frac s{\ln(\mu_0/\Lambda_0)}\right)^{21/5},$
$b=5/(12\pi^2)$ and $Y_0=(u(\mu_0)-1)/(3u(\mu_0)+4)$.

\section{Thermodynamics of the tricritical point}
\label{sec:thermo}

The one-loop study of the tricritical point in the $SU(2)_L\times
SU(2)_R$ linear sigma model was done in \cite{Jakovac:2003ar} using an
expansion in the number of flavors. We quote below the equations (12)
and (14) of that work which determine the position of the tricritical
point in the $\mu_q-T$ plane. Using the present notation for the
couplings these equations read:
\begin{eqnarray}
  && m_\ph^2 +T^2\left[\frac\lambda3+g^2 +\frac{3g^2}{\pi^2}
    \alpha^2\right] =0,\nn
  && \lambda + \frac{3g^4}{\pi^2}\big[ \ln(\beta\mu)
    - {\cal F}(\alpha) \big]=0,
\end{eqnarray}
where $\beta=1/T$ is the inverse temperature and $\alpha=\beta \mu_q$
with $\mu_q$ the quark baryon chemical potential. The function 
${\cal F}$ reads
\begin{equation}
  {\cal F}(\alpha) = 1 -\gamma_E + \ln 2 -\frac\d{\d s}
  \left[\mathrm{Li}_s(-e^\alpha) + \mathrm{Li}_s(-e^{-\alpha})
    \right]\Big|_{s=0}.
\end{equation}
This is a monotonously increasing function of its argument, ${\cal
  F}(0)= 1.5675$.

We choose the scale $\mu=e^\xi T$, where $\xi$ is a number of ${\cal
  O}(1)$. Then the logarithm yields $\xi$, which effectively modifies
${\cal F}\to \bar{\cal F} = {\cal F} - \xi$. In case of spontaneous
symmetry breaking $m_\varphi^2<0$, and it is useful to rescale all the
masses with the tree level sigma mass $m_\sigma^2 = -2m_\varphi^2$ at scale
$\mu$. Then the equations to solve will be
\begin{equation}
  \label{BT}
  \frac12 = T^2 g^2(\mu)\left[\frac{u(\mu)}{6} +1 + \frac{3}{\pi^2}
    \alpha^2\right],\qquad u(\mu) = \frac{6}{\pi^2}
  g^2(\mu)\,\bar{\cal F}(\alpha). 
\end{equation}

In the complete problem therefore there are 5+2 parameters: at the UV
scale $m_s$ we have $m_s,\, m_a,\, g^2,\, u, \,u_2$, and also we have
$\mu$ and $T$ at the IR scale. The light mass $m_\varphi$ or the
corresponding $m_{\sigma}$ is used as a mass unit. At the TCP
$\alpha=\beta \mu_q$ and $T$ can be determined as functions of the
UV parameters:
\begin{equation}
  {\cal G}_0 : m_s,m_a,g^2,u,u_2 \mapsto \alpha_c, T_c.
\end{equation}

The final output of the investigation should be, of course
$\alpha_c(m_s)$ and $T_c(m_s)$. But as we just have seen, even in the
chiral limit of $m_{ud}=0$ the problem is five dimensional instead of
one dimensional. For a sensible prediction we have to say something
about the strange mass dependence of $m_a,g^2,u,u_2$ -- these
functions should come from the underlying theory, now QCD. Since we
do not have this information, we have to assume something sensible.

In the light of the previous subsections we make some
approximations: we can fix $u_2(m_s)=2$ and for $u$ we consider two cases:
$u(m_s)=4$ and $u(m_s)=10.$ 
The remaining function 
\begin{equation}
  {\cal G} : m_s,m_a,g^2 \mapsto \alpha_c, T_c
\end{equation}
can be plotted as shown in Fig.~\ref{fig:surface}. The detailed
numerical strategy to solve the system and obtain this plot is given
in Appendix \ref{app:numstrat}. Fig.~\ref{fig:surface} shows surfaces
in the $m_s,m_a,g^2$ parameter space leading to some fixed value of
$\alpha_c.$ The $\mu_q/T=0$ critical surface is a limiting one, in
the sense that surfaces with $\mu_q/T>0$ all lie on one of its side,
they never cross each other. Moreover the normal vector of the surface
pointing to positive $\mu_q/T$ always have \emph{negative} $m_s$
component -- in this sense we can say that going on the direction of
the largest $\mu_q/T$ change, the surfaces bend downwards in $m_s$.

In order to show that different values of $u$ do not change
qualitatively the result we plot the $\mu_q/T=0.8$ surface obtained using
$u(m_s)=10$ and $u(m_s)=6$, and rescale the $g^2$ values of the latter
by a factor of 1.5. The two surfaces can be seen on the right panel of
Fig.~\ref{fig:surface}. As the plot shows, the surfaces have the same
characteristics. 

\begin{figure}[!t]
  \centering
  \includegraphics[keepaspectratio,width=0.495\textwidth,angle=0]{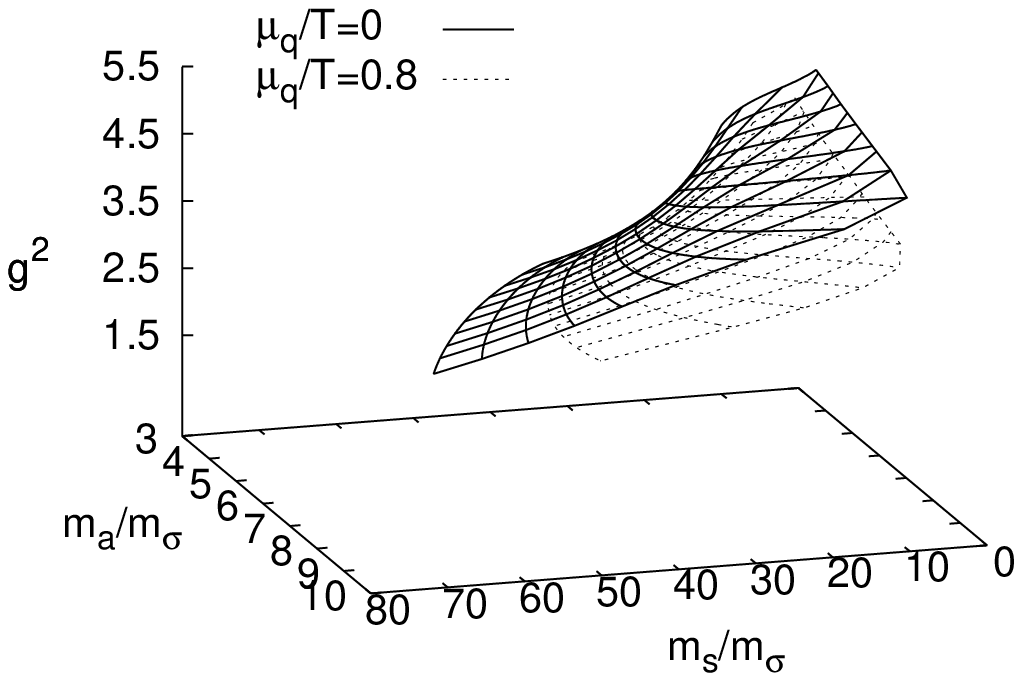}
  \includegraphics[keepaspectratio,width=0.495\textwidth,angle=0]{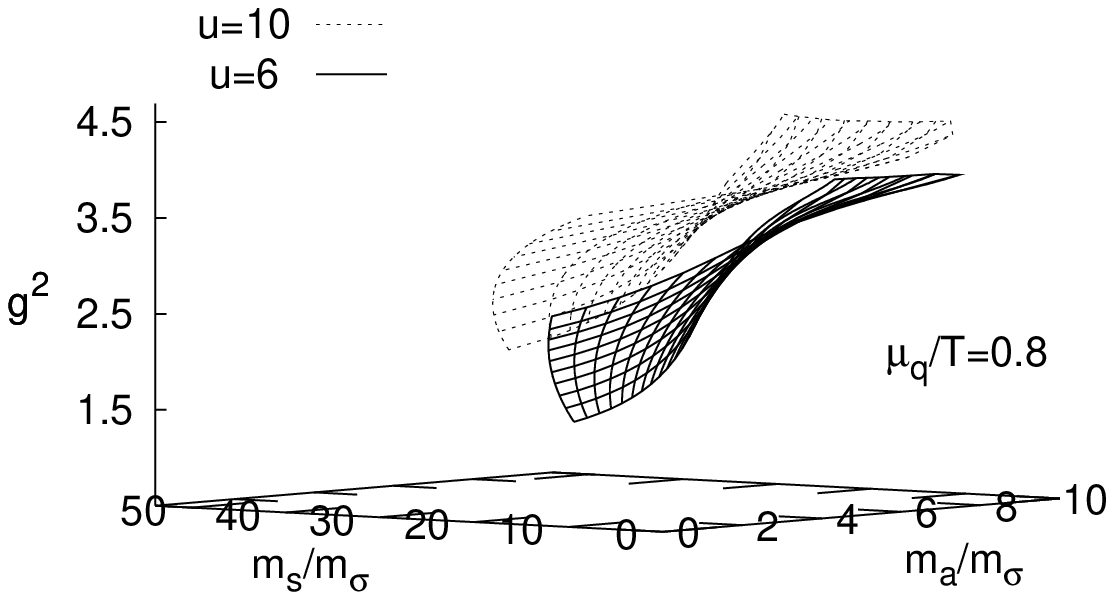}
  \caption{ The dependence of $\alpha_c$ ($\mu_q/T$ at the TCP) on the 
    parameters for $u_2(m_s)=2$. The left panel is obtained with $u(m_s)=10$.
    The value of $\Lambda_0$ is fix along the lines of the surfaces 
    directed towards the origin of the $m_s-m_a$ plane. For further 
    information see the main text.
}
  \label{fig:surface}
\end{figure}

\subsection{Implication of the results for QCD}

In real QCD we cannot change $m_a,\, m_s$ and $g^2$ independently. If
we knew the $m_s$ dependence of $m_a$ and $g^2$ then we would have a
curve in the $m_s-m_a-g^2$ space parametrized by $m_s$. This line
would go through the critical surfaces characterized by fix $\mu_q/T$,
and then we could determine the $m_s(\mu_q/T)$ function. Since we do
not have any information on the $m_s$ dependence of the parameters we
explore several possibilities by fixing the value of one of the
parameters.

\begin{figure}[!t]
  \centering
  \includegraphics[keepaspectratio,width=0.495\textwidth,angle=0]{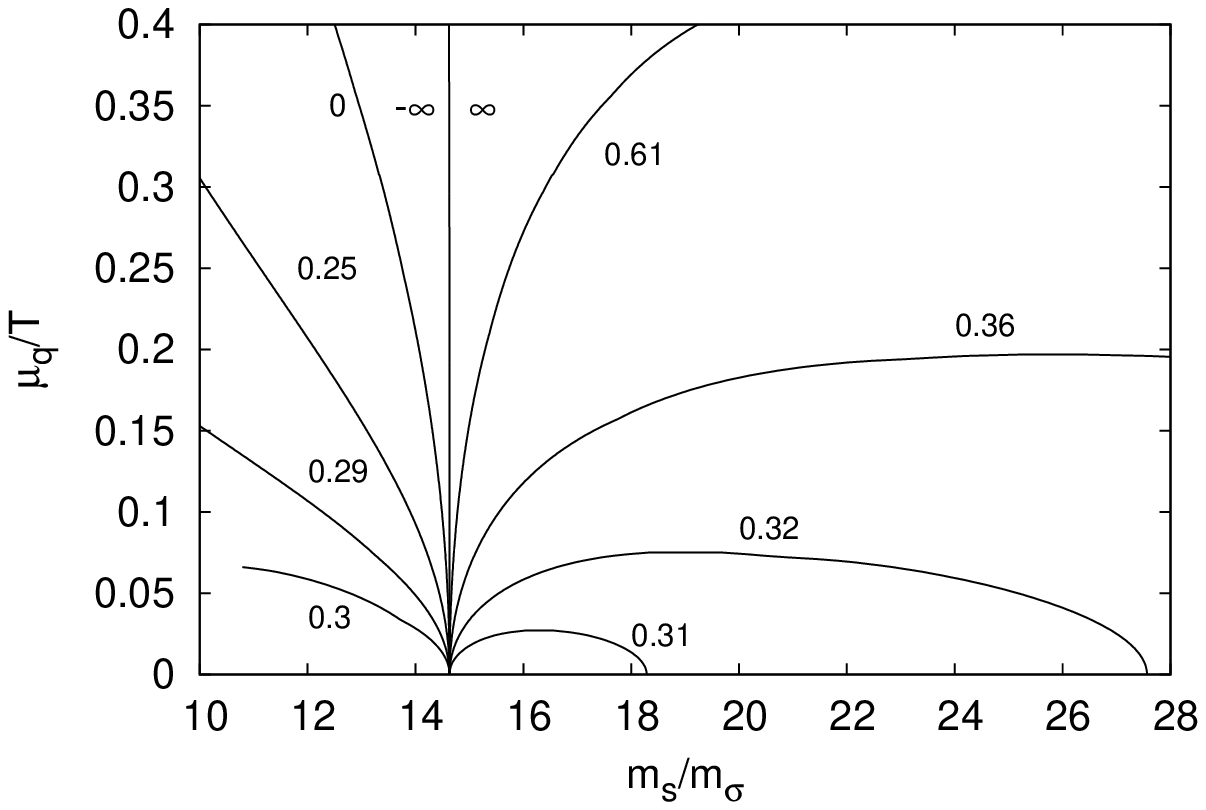}
  \includegraphics[keepaspectratio,width=0.495\textwidth,angle=0]{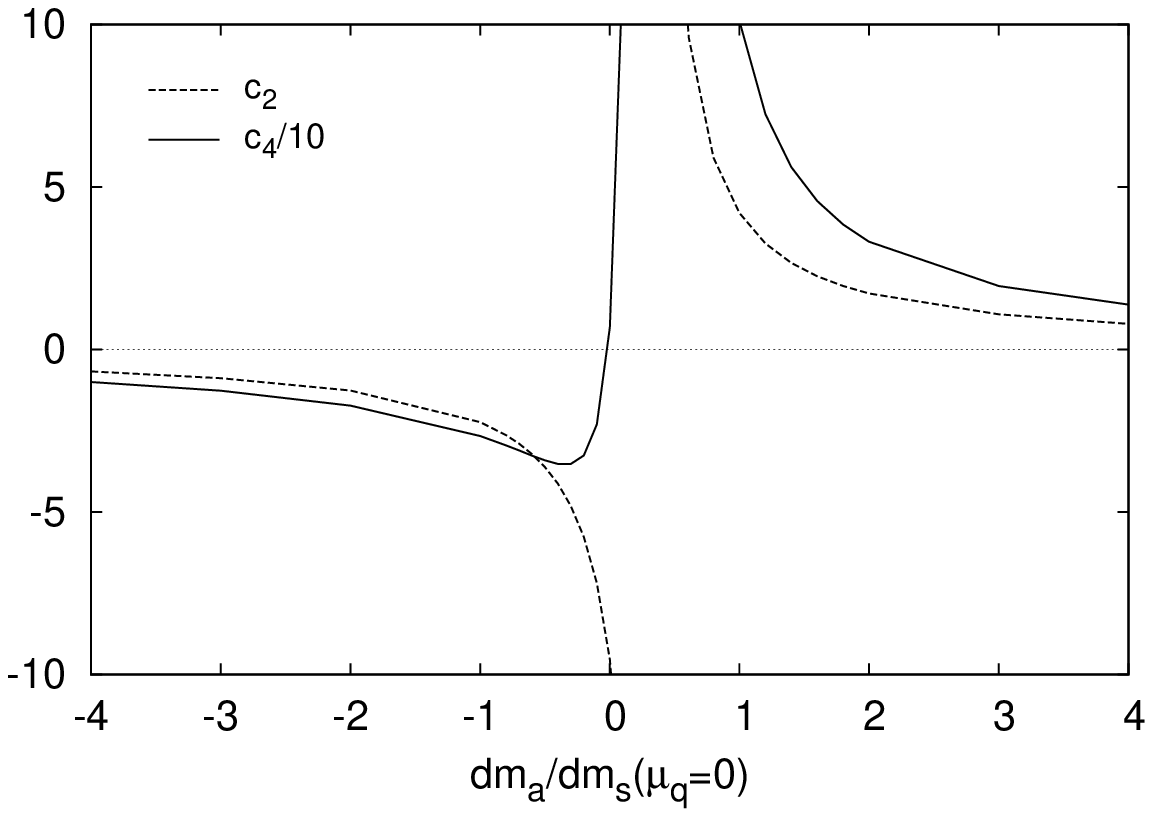}
  \caption{Left panel: The $m_s$-dependence of the critical $\mu_q/T$ obtained 
    for $u=10$ and $g^2=3.$ The labels on the curves indicate the value of 
    $d m_a/(d m_s)(\mu=0).$ Right panel: The first two coefficients of the 
    Taylor expansion in Eq.~(\ref{Eq:ms_expansion}).}
  \label{fig:dat1}
\end{figure}

For a constant value of $g^2$ the $m_s$-dependence of the critical
$\mu_q/T$ is shown in Fig.~\ref{fig:dat1}.  One can see that the
behavior of this curve depends strongly on how $m_a$ depends on
$m_s$. Characterized by $dm_a/dm_s$ at $\mu_q=0$, there is a
limiting value, and tricritical curves with smaller value of
$dm_a/dm_s$ bend downwards (negative curvature), for larger values
they bend upwards (positive curvature). 

The standard characterization of the behavior of $m_s(\mu_q)$ near 
$\mu_q=0$ is through the Taylor series
\cite{deForcrand:2010ys}:
\begin{equation}
  \frac{m_s(\mu_q)}{m_s(0)} = 1 + \sum\limits_{k=1} c_k
  \left(\frac{\mu_q}{\pi T}\right)^{2 k}.
\label{Eq:ms_expansion}
\end{equation}
The first two nontrivial terms $c_2$ and $c_4$ are shown in Fig.~\ref{fig:dat1}
in case of a constant $g^2$. The singularity corresponds to that value of the 
$m_s$ for which the curvature changes sign. It is remarkable that by
changing continuously from negative to positive curvatures ($c_2$
values) we have to go through a singularity.
\begin{figure}[htbp]
  \centering
  \includegraphics[keepaspectratio,width=0.5\textwidth,angle=0]{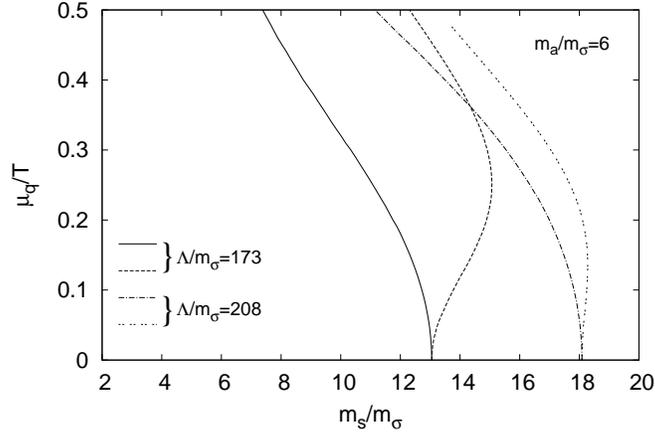}
  \caption{Effect of a $\mu$-dependent $m_a$ on the $m_s$-dependence
    of the critical $\mu_q/T$ obtained for $u=10.$ From the two pairs
    of lines having the same values of $\Lambda/m_\sigma$ and
    $m_a/m_\sigma,$ the one at the right is obtained by solving
    (\ref{BT}) with a $\mu$-dependent $m_a.$}
  \label{fig:dat2}
\end{figure}

To have a hint on which curve could be the physical one we recall that
the anomaly is mainly a gauge effect, connected to the presence of the
instantons \cite{'tHooft:1976fv, 'tHooft:1976up}. This suggests that
The dependence of $m_a$ on $m_s$ should be quite small, so the
physical line is near to $dm_a/dm_s=0$. Note that all tricritical
surfaces in this model with $m_a=$constant bend \emph{downwards}.

We can compare our results with the numerical findings of
\cite{deForcrand:2010ys}. They found $c_2=-3.3(3)$, $c_4=-47(20)$ with
$N_f=3$ degenerate quarks. The $c_2$ curve on Fig.~\ref{fig:dat1}
reproduces this value at $dm_a/dm_s=-0.58$ with a corresponding value
of $c_4=-33.0$. This is in the same order of magnitude as in the MC
simulation.

In Ref.~\cite{Chen:2009gv} it was shown that if the strength of the
$U(1)$ anomaly, parameter $C$ in the Lagrangian (\ref{SUNSUNLag}), is
made $\mu_q$-dependent then the critical surface can have a
nonmonotonic shape.  Since $C$ influences $m_a,$ we can observe the
same effect by considering the following dependence of $m_a$ on
$\mu_q$, when solving (\ref{BT}):
\begin{equation}
\frac{m_a(\mu_q)}{m_\sigma}=1+\left(\frac{m_a(0)}{m_\sigma}-1\right)
e^{-\mu_q^2/\mu_{q,0}^2}.
\end{equation}
For most of the curves with this chemical potential dependent anomaly
are very similar to what is obtained earlier, however one can observe
more exotic behavior, too, for some cases. For
$\mu_{q,0}/m_\sigma=0.17$ the result is shown in Fig.~\ref{fig:dat2}:
here the surface starts to bend upwards, and later it turns back. This
behavior can be very mild (as in our example with
$\Lambda/m_\sigma=208$). In this case coarse lattice measurement would
only detect the negative curvature, a high precision lattice
measurement is necessary to reveal the positive curvature near
$\mu_q=0$. It is interesting that, considering the global behavior of
the curve, the coarser lattice would give a more reliable result in
this case.

\section{Conclusions}
\label{sec:conc}

We discussed the behavior of the line of tricritical points (TCP) in
the chiral limit ($m_{ud}=0$) of the $U(3)_L\times U(3)_R$ quark
model. We assumed that the value of the strange quark mass, where the
TCP hits the $\mu=0$ line is much larger than the critical temperature
$T_c$. This is a good approximation in QCD, where the critical
temperature is of order 160~MeV, while the constituent strange quark
mass is about 450~MeV already at the physical point, and we expect
that the second order line reaches the chiral line ($m_{ud}=0$) at
much higher $m_s$ masses. We also assumed that the $\eta-a_0$ meson
sector, which is heavier than the $\sigma-\pi$ sector because of the
anomaly, is also much heavier than $T_c$. This is again plausible,
since already at the physical point $m_a\sim1$~GeV.

Under these circumstances the strange and the $\eta-a_0$ sector
decouples from the point of view of the thermodynamics, which is
completely determined by the light degrees of freedom, the
$\sigma-\pi$ sector. The only way how the heavy sector can influence
the thermodynamics is through the values of the parameters of the
effective theory. To achieve this goal we have to follow the running
of the different parameters as well as the degrees of freedom from the
heavy scales down to the thermodynamic scale. This can be performed by
following the renormalization group (RG) flow with given degrees of
freedom, and determine the parameters of the effective theories by
matching when the degrees of freedom change. The former yield
logarithmic dependence on the heavy scale, the latter effect is
power-suppressed. Therefore in this work the RG flow is determined at
one-loop level, and the matching is kept at tree level.

With the decoupling of the strange and $\eta-a_0$ meson sector,
respectively, there are two stages of effective models in the
$U(3)_L\times U(3)_R$ linear sigma model: the first is the
$U(2)_L\times U(2)_R,$ while the second is the $SU(2)_L\times SU(2)_R$
linear sigma model. We determined the corresponding beta functions in
these models, and solved the RG flow down to the scale of the
temperature $T.$ The thermodynamics is determined at one-loop level in
\cite{Jakovac:2003ar} -- we now included the running coupling constant
in the result.

As a result we can determine the free energy for any given parameter
sets, and we can determine those points where tricritical points (TCP)
are located. With some plausible assumptions, the TCP's with fixed
$\alpha=\mu_q/T$ (where $\mu_q$ is the quark chemical potential) form
a surface in the $m_s-m_a-g^2$ space in the $U(3)_L\times U(3)_R$
linear sigma model. Surfaces for different $\alpha$ never cross,
therefore the $\alpha=0$ surface is a limiting surface.

If we want to draw consequences for QCD, we have to specify how $m_a$
and $g^2$ depend on $m_s$ in the chiral $u,d$ regime. Since it is not
known, we explored several possibilities. Depending on the details,
this model can describe an upward bending (positive curvature) surface
or a downward bending (negative curvature) surface. Taking into
account the explicit chemical potential dependence of the anomaly
constant, the curvature we can change from positive to negative
curvatures along the curve. In this case the value of the curvature at
$\mu_q=0$ would yield false information about the global behavior of
the curve.

\begin{acknowledgments}
  The authors benefited from discussions with Andr\'as Patk\'os and
  P\'eter Petreczky. This work is supported by the Hungarian Research
  Fund (OTKA) under contract Nos. T068108 and K77534.  
\end{acknowledgments}

\appendix


\section{Renormalization group equations in the $U(2)_L\times U(2)_R$ model}
\label{sec:U2}

We start from the renormalized Lagrangian of \eqref{L2c}, and introduce
the counterterm Lagrangian which in Fourier space reads
\begin{equation}
  \delta {\cal L}_{U(2)} =\delta Z_\psi \bar \psi_2
  \slash\hspace*{-0.5em}k\psi_2 - \delta g\bar\psi_2 (\ph_5-i\gamma_5
  a_5)\psi_2 + \frac{\delta Z}2 \ph(k^2-\delta m_\ph^2)\ph +
  \frac{\delta Z}2 a(k^2-\delta m_a^2)a - \frac{\delta
    \lambda}4(\ph^2+a^2)^2 -\frac{\delta \lambda_2}2(\ph^2a^2-(\ph a)^2),
\label{Eq:L2_counter}
\end{equation}
where we used the shorthand $\delta m_{\ph/a}^2 = \delta m^2\mp \delta c$
and the observation that the wave function renormalization
for the $\ph$ and $a$ sector is the same. 

The goal is to determine the counterterms at one-loop level. To this
end we work at zero temperature in the symmetric phase. The method is
to determine the expectation value of some physical observables and
require finiteness.  In order to simplify the treatment we introduce a
background field for the $\sigma$ field through the shift
$\sigma\to\sigma+x$. The expansion in $x$ is used for zero momentum
external legs. The new Lagrangian is obtained from \eqref{L2c} reads
\begin{eqnarray}
  {\cal L}_{U(2)} =&& -\frac{m_\ph^2}2 x^2 - \frac\lambda4 x^4 - \sigma
  x(m_\ph^2 + \lambda x^2) + \bar\psi_2(i\dslash-m_\psi)\psi_2 +
  \frac12(\d_\mu\ph)^2 + \frac12(\d_\mu a)^2 - \frac{m_\sigma^2}2
  \sigma^2 - \frac{m_\pi^2}2\pi_i^2 -\frac{m_\eta^2}2\eta^2 -
  \frac{m_A^2}2 a_i^2\nn&& - g\bar\psi_2(\ph_5-i\gamma_5a_5)\psi_2
  -\lambda x \sigma(\ph^2+a^2) -\lambda_2 x\sigma a_i^2 - \lambda_2 x
  \eta\pi_i a_i - \frac\lambda4(\ph^2+a^2)^2 - \frac{\lambda_2}2
  \big[\ph^2 a^2 - (\ph a)^2\big],
\end{eqnarray}
where
\begin{equation}
  m_\psi = gx,\qquad
  m_\sigma^2 = m_\ph^2 + 3\lambda x^2,\qquad
  m_\pi^2 = m_\ph^2 + \lambda x^2,\qquad
  m_\eta^2 = m_a^2 + \lambda x^2,\qquad
  m_A^2 = m_a^2 + \lambda x^2 + \lambda_2 x^2,
\end{equation}
with $m_\varphi^2$ and $m_a^2$ defined below (\ref{L2c}).

\subsection{The fermionic wave function and $g$ renormalization}

We calculate on the $x$ background the fermion self-energy
$\Sigma_\psi = i\exv{\T\psi_2\bar\psi_2}_{amp}$. Introducing the notation
\hbox{$\displaystyle \pint4p=\int\frac{d^4 p}{(2\pi)^4}$} and using standard
Feynman rules we find
\begin{equation}
  \Sigma_\psi(k) = -\delta Z_\psi \slash\hspace*{-0.5em}k +x \delta g-
  ig^2 \pint4p \Big[ \big(iG_\sigma(p-k)+3iG_a(p-k)\big)i{\cal G}(p) -
    \big(iG_\eta(p-k)+3iG_\pi(p-k)\big) \gamma_5 i{\cal G}(p)\gamma_5 \Big],
\end{equation}
where $G$ and ${\cal G}$ are the bosonic and fermionic propagators defined as
\begin{equation}
  G(p) = \frac1{p^2-m^2+i\ep},\qquad {\cal G}(p) =
  \frac{\slash\hspace*{-0.5em}p +m}{p^2-m^2 +i\ep},
\label{Eq:propagators}
\end{equation}
with the corresponding masses.

Since the integral has mass dimension, the part proportional to
$\slash\hspace*{-0.5em}k$ or $x$ are dimensionless which means that
they are at most logarithmically divergent. Therefore the masses
should be taken into account only through a Taylor expansion. But, the
expansion in the bosonic masses yields $m^2\sim x^2$ terms, which are
convergent, so we can forget about the bosonic masses. The same is
true for the fermionic mass in the denominator. What remains for the
divergent piece is
\begin{equation}
  \Sigma_\psi^{\mathrm{div}}(k) = -\delta Z_\psi \slash\hspace*{-0.5em}k
  +x \delta g + 4ig^2 \pint4p G(p-k)\big[{\cal G}(p) - \gamma_5{\cal
      G}(p)\gamma_5 \big].
\end{equation}
In the numerator we find
$ \slash\hspace*{-0.5em}p +m - \gamma_5(\slash\hspace*{-0.5em}p
  +m)\gamma_s = 2\slash\hspace*{-0.5em}p,
$
which results in
\begin{equation}
  \delta g=0.
\end{equation}
For $\delta Z_\psi$ we have to calculate the remaining integral. Doing this 
with standard techniques (cf. for example \cite{PeskinSchroeder}) using
cut-off regularization we find
\begin{equation}
  \Sigma_\psi^{\mathrm{div}}(k) =\left(-\delta Z_\psi -
    \frac{g^2}{4\pi^2}\ln\frac{\Lambda^2}{\mu^2}\right)\slash\hspace*{-0.5em}k,
\end{equation}
and so, the counterterm ensuring the finiteness of  $\Sigma_\psi(k)$ is
\begin{equation}
\delta Z_\psi =- \frac{g^2}{4\pi^2} \ln\frac{\Lambda^2}{\mu^2}.
\end{equation}

\subsection{The bosonic wave function and $\lambda$ renormalization}

We calculate next the $\sigma$ self-energy on the given $x$
background. We find
\begin{eqnarray}
  \Sigma_\sigma(k) &=&-\delta Z k^2+\delta m_\ph^2+3\delta\lambda x^2
  + 3\lambda \pint4p iG_\sigma(p) + 3\lambda \pint4p iG_\pi(p) +
  \lambda \pint4p iG_\eta(p)\nn&& + 3(\lambda+\lambda_2) \pint4p
  iG_a(p) - ig^2 \pint4p\, \Tr\left[{\cal G}(p-k){\cal G}(p)\right]
  +18i\lambda^2 x^2 \pint4p G_\sigma(p-k) G_\sigma(p)\nn&& 
  +6i\lambda^2 x^2 \pint4p G_\pi(p-k) G_\pi(p)
  +2i\lambda^2 x^2 \pint4p G_\eta(p-k) G_\eta(p)
  +6i(\lambda+\lambda_2)^2 x^2 \pint4p G_a(p-k) G_a(p).\hspace*{1em}
\end{eqnarray}
The minus sign is because the fermionic bubble involves a closed
fermion loop.  To determine $\delta Z$ and $\delta m_\ph$ we need only
the $x=0$ sector:
\begin{eqnarray}
  \Sigma_\sigma(k,x=0) = -\delta Z k^2 + \delta m_\ph^2 + 6\lambda \pint4p
  iG_\ph(p) + (4\lambda+3\lambda_2) \pint4p iG_a(p) - ig^2 \pint4p\,
  \Tr\big[{\cal G}(p-k){\cal G}(p)\big].\hspace*{2em}
\end{eqnarray}
After evaluating the integrals, without writing the $\Lambda^2$ corrections 
we find
\begin{equation}
  \Sigma_{\sigma}^{\mathrm{div}}(k,x=0) = -\delta Z k^2 +\delta m_\ph^2
  +\frac{6\lambda}{16\pi^2}\,m_\ph^2 \ln\frac{m_\ph^2}{\Lambda^2}
  +\frac{4\lambda+3\lambda_2}{16\pi^2}\,m_a^2 \ln\frac{m_a^2}{\Lambda^2}
  - \frac{g^2}{24\pi^2} \, k^2 \ln\frac{k^2}{\Lambda^2}.
\end{equation}
Therefore
\begin{equation}
  \delta Z = \frac{g^2}{12\pi^2} \,
  \ln\frac{\Lambda^2}{\mu^2},\qquad \delta m_\ph^2 =
  \frac{1}{16\pi^2} \ln\frac{\Lambda^2}{\mu^2}
  \big[ 6 \lambda m_\ph^2 +(4\lambda+3\lambda_2)\,m_a^2 \big].
\label{Eq:dZ_dm2_phi}
\end{equation}

For the determination of $\delta\lambda$ we need the self-energy at $k=0$. 
After evaluating the integrals we find
\begin{eqnarray}
  \Sigma_{\sigma}^{\mathrm{div}}(k=0) &&= \delta m_\ph^2 - \frac1{16\pi^2}
  \ln\frac{\Lambda^2}{\mu^2} \left[ 6\lambda
    m_\ph^2+(4\lambda+3\lambda_2) m_a^2 \right]
  +3\delta\lambda x^2 - \frac{3x^2}{16\pi^2}
  \ln\frac{\Lambda^2}{\mu^2} \left[ 13\lambda^2 +
    3(\lambda+\lambda_2)^2 - 4g^4\right].\qquad
\end{eqnarray}
We obtain for $\delta m^2_\ph$ the previous result given in
(\ref{Eq:dZ_dm2_phi}), and we also have
\begin{equation}
  \label{deltalambda}
  \delta\lambda = \frac1{16\pi^2} \ln\frac{\Lambda^2}{\mu^2} \left[
    13\lambda^2 + 3(\lambda+\lambda_2)^2 - 4g^4\right].
\end{equation}

\subsection{Renormalization of $\lambda_2$}

We apply the procedure above, but now for the $a$ self-energy. Since
the wave function renormalization is the same as for $\ph$ we need
only the $k=0$ case. We find:
\begin{eqnarray}
 \Sigma_a(k=0) &=& \delta m_a^2+(\delta\lambda+\delta\lambda_2)x^2 +
  5\lambda \pint4p iG_a(p) + (3\lambda+2\lambda_2) \pint4p
  iG_\pi(p)\nn&& + \lambda \pint4p iG_\eta(p) + (\lambda+\lambda_2)
  \pint4p iG_\sigma(p) - ig^2 \pint4p\, \Tr\big[{\cal G}(p){\cal
      G}(p)\big]\nn&& +4i(\lambda+\lambda_2)^2 x^2 \pint4p G_\sigma(p)
  G_a(p) +i\lambda_2^2 x^2 \pint4p G_\pi(p) G_\eta(p).
\end{eqnarray}
After evaluating the integrals we find for the divergent pieces
\begin{eqnarray}
  \Sigma_{a}^{\mathrm{div}}(k=0) &&= \delta m_a^2+(\delta\lambda+\delta\lambda_2)x^2 -
  \frac1{16\pi^2} \ln\frac{\Lambda^2}{\mu^2} \left[ 6\lambda
    m_a^2+(4\lambda+3\lambda_2) m_\ph^2 \right]- \frac{x^2}{16\pi^2}
  \ln\frac{\Lambda^2}{\mu^2} \big[ 16\lambda^2 +18\lambda\lambda_2 +
    5\lambda_2^2 - 12g^4\big].\nn
\label{Eq:sigma_a_div}
\end{eqnarray}
For the $a$-mass counterterm we find he following expression
\begin{equation}
   \delta m_a^2 = \frac1{16\pi^2} \ln\frac{\Lambda^2}{\mu^2} \big[
     6\lambda m_a^2 +(4\lambda+3\lambda_2)m_\ph^2\big],
\end{equation}
which is the same as the expression for the $\sigma$ mass, with the
$m_\ph\leftrightarrow m_a$ interchange. This shows that the sum and
the difference of the masses are renormalized multiplicatively: since
$m_{\ph/a}^2 = m^2\mp c$, so we find
\begin{equation}
  \delta c = c\,\frac{2\lambda-3\lambda_2}{16\pi^2}
  \ln\frac{\Lambda^2}{\mu^2},\qquad \delta m^2 = m^2\,
  \frac{10\lambda+3\lambda_2}{16\pi^2} \ln\frac{\Lambda^2}{\mu^2}.
\end{equation}
From (\ref{Eq:sigma_a_div}) we can also read off the counterterm for 
$\lambda+\lambda_2$:
\begin{equation}
  \delta \lambda+\delta\lambda_2 = \frac1{16\pi^2}
  \ln\frac{\Lambda^2}{\mu^2} \big[ 16\lambda^2 +18\lambda\lambda_2 +
    5\lambda_2^2 - 12g^4\big].
\end{equation}
Comparing it with the expression of $\delta\lambda$ given in
\eqref{deltalambda} we find
\begin{equation}
  \delta\lambda_2 = \frac2{16\pi^2} \ln\frac{\Lambda^2}{\mu^2} \big[
    \lambda_2(6\lambda+\lambda_2) -4g^4\big].
\end{equation}

\subsection{$\beta$ functions}

The bare-field Lagrangian reads
\begin{equation}
  {\cal L}_{U(2),0} =\bar \psi_{02} \bigl[i\dslash -
  g_0(\ph_{05}-i\gamma_5a_{05})\bigr]\psi_{02} + \frac12 (\d_\mu\ph_0)^2
  + \frac12 (\d_\mu a_0)^2 -\frac{m_{0\ph}^2}2\ph_0^2 - \frac{m_{0a}^2}2a_0^2 -
  \frac{\lambda_0}4(\ph_0^2+a_0^2)^2
  -\frac{\lambda_{02}}2\big[\ph_0^2a_0^2-(\ph_0 a_0)^2\big],
\end{equation}
where all the fields and couplings are bare. The bare couplings are RG
invariant, since they depend only on the regularization: 
\begin{equation}
  \frac{d g_0}{d\ln\mu} = \frac{d m_{0\ph}^2}{d\ln\mu} =  \frac{d
    m_{0a}^2}{d\ln\mu} = \frac{d\lambda_0}{d\ln\mu} =
  \frac{d\lambda_{02}}{d\ln\mu} = 0,
\end{equation}
where
\begin{equation}
  \frac d{d\ln\mu} =\frac\d{\d\ln\mu} + \beta_g \frac{\d}{\d g} +
  \beta_\lambda \frac{\d}{\d\lambda} + \beta_2 \frac{\d}{\d\lambda_2}
  +\gamma_\ph \frac{\d}{\d m_\ph^2} +\gamma_a \frac{\d}{\d m_a^2}.
\label{Eq:deriv_chain}
\end{equation}

To obtain the bare quantities from the counterterms, we first have to
change to renormalized fields $\psi_{02}=Z_\psi^{1/2} \psi_2$,
$\ph_0=Z^{1/2}\ph,$ and $a_0=Z^{1/2}a$:
\begin{eqnarray}
  {\cal L}_2 &=&\bar \psi_2 \bigl[Z_\psi i\dslash - Z_\psi Z^{1/2}
  g_0(\ph_5 - i\gamma_5a_5)\bigr]\psi_2 + \frac Z2 (\d_\mu\ph)^2 + \frac
  Z2 (\d_\mu a)^2 \nn &&-\frac{Zm_{0\ph}^2}2\ph^2 - \frac{Zm_{0a}^2}2a^2 -
  \frac{Z^2\lambda_0}4(\ph^2+a^2)^2
  -\frac{Z^2\lambda_{02}}2\big[\ph^2a^2-(\ph a)^2\big].
\end{eqnarray}
Comparing it with the renormalized Lagrangian defined as the sum of \eqref{L2c} and (\ref{Eq:L2_counter}) we find $Z_\psi = 1+\delta Z_\psi$ and 
$Z = 1+\delta Z,$ so that the relations between the bare couplings and 
counterterms read
\begin{eqnarray}
  && Z_\psi Z^{1/2} g_0 = g+\delta g,\quad 
  Zm_{0\ph}^2=m_\ph^2+\delta m_\ph^2,\quad
  Zm_{0a}^2=m_a^2+\delta m_a^2,\nn
  && Z^2\lambda_0=\lambda+\delta\lambda,\quad
  Z^2\lambda_{02}=\lambda_2+\delta\lambda_2.
\end{eqnarray}
These relations can be inverted, and at one-loop level we obtain:
\begin{eqnarray}
  && g_0=g -\left(\frac12\delta Z+\delta Z_\psi\right)g + \delta g,\qquad
  m_{0\ph}^2=m_\ph^2- \delta Z m_\ph^2+ \delta m_\ph^2 ,\qquad
  m_{0a}^2=m_a^2- \delta Z m_a^2+ \delta m_a^2 ,\nn&&
  \lambda_0=\lambda-2\delta Z\lambda+\delta\lambda,\qquad
  \lambda_{02}=\lambda_2-2\delta Z\lambda_2+\delta\lambda_2.
\end{eqnarray}
Perturbative hierarchy requires that when there is a $\ln\mu$
dependence in the quantity, then only the $\d/(\d\ln\mu)$ derivative
acts on it. Then, using (\ref{Eq:deriv_chain}) we find
\begin{equation}
  \frac{dg_0}{d\ln\mu} = \beta_g - \frac{\d}{\d\ln\mu}\left[ \left(\frac12
    \delta Z+\delta Z_\psi\right)g - \delta g \right] =0,
\end{equation}
and in consequence
\begin{equation}
   \beta_g = \frac{\d}{\d\ln\mu}\left[ \left(\frac12
    \delta Z+\delta Z_\psi\right)g - \delta g \right].
\label{Eq:beta_a}
\end{equation}
In a similar way we find
\begin{eqnarray}
  && \gamma_\ph = \frac{\d}{\d\ln\mu}\left[ \delta Z m_\ph^2 - \delta
    m_\ph^2\right],\qquad \gamma_a = \frac{\d}{\d\ln\mu}\big[ \delta
    Z m_a^2 - \delta m_a^2\big],\nn&&
  \beta_\lambda = \frac{\d}{\d\ln\mu}\left[2\delta Z \lambda - \delta
    \lambda\right],\qquad
  \beta_2 = \frac{\d}{\d\ln\mu}\left[2\delta Z \lambda_2 - \delta
    \lambda_2\right]. 
\label{Eq:beta_b}
\end{eqnarray}
Using the expression of the counterterms determined in previous
subsections of this sections we have:
\begin{eqnarray}
  &&\frac{\d \delta Z_\psi}{\d\ln\mu} = \frac{g^2}{2\pi^2},\qquad
  \frac{\d \delta Z}{\d\ln\mu} = -\frac{g^2}{6\pi^2},\quad
  \frac{\d\delta g}{\d\ln\mu} =0,\nn&&
  \frac{\d\delta\lambda}{\d\ln\mu} = -\frac{1}{8\pi^2} \left[
    13\lambda^2 + 3(\lambda+\lambda_2)^2 - 4g^4\right],\qquad
  \frac{\d\delta\lambda_2}{\d\ln\mu} = -\frac{1}{4\pi^2} \left[
    \lambda_2(6\lambda+\lambda_2) -4g^4\right],\nn&&
  \frac{\d \delta m_{\ph/a}^2}{\d\ln\mu} = - \frac{1}{8\pi^2} \left[
     6\lambda m_{\ph/a}^2 +(4\lambda+3\lambda_2)m_{a/\ph}^2\right].
\end{eqnarray} 
With these expressions, we obtain from (\ref{Eq:beta_a}) and 
(\ref{Eq:beta_b}) the following one-loop $\beta$-functions:
\begin{subequations}
\begin{eqnarray}
  \label{Eq:U2_g_run}
  &&\frac{dg}{d\ln\mu}= \beta_g=\frac{5g^3}{12\pi^2},\\
  \label{Eq:lambda_run}
  &&\frac{d\lambda}{d\ln\mu} =\beta_\lambda = \frac1{8\pi^2}
  \left[13\lambda^2 + 3(\lambda+\lambda_2)^2 - 4g^4-\frac83
    g^2\lambda\right],\\
  \label{Eq:lambda2_run}
  &&\frac{d\lambda_2}{d\ln\mu}= \beta_2 = \frac1{4\pi^2} \left[
    \lambda_2(6\lambda+\lambda_2) -4g^4-\frac43
    g^2\lambda_2\right].\\
  &&\frac{dm^2_{\ph/a}}{d\ln\mu} = \gamma_{\ph/a} = \frac1{8\pi^2}
  \left[ \Big(6\lambda-\frac43g^2\Big) m_{\ph/a}^2 +
    (4\lambda+3\lambda_2)m_{a/\ph}^2\right].
\end{eqnarray}
\label{U2beta}
\end{subequations}

\section{Renormalization group in the  $SU(2)_L\times SU(2)_R$ linear sigma model}
\label{sec:O4}

We start from the renormalized Lagrangian \eqref{LO4} and add to it the
the following counterterm Lagrangian:
\begin{equation}
  \delta {\cal L}_{SU(2)} =\bar \psi \bigl[Z_\psi i\dslash -\delta
  g(\sigma+i\tau_i\pi_i)\bigr]\psi + \frac Z2 (\d_\mu\ph)^2 
  -\frac{\delta m^2_\varphi}2\ph^2 - \frac{\delta\lambda}4(\sigma^2+\pi_i^2)^2.
\end{equation}
The goal is to determine the counterterms at one-loop level. To do
this, we will follow the same strategy as in Appendix~\ref{sec:U2}. To
facilitate the discussion we introduce again the  background field $x.$
After the shift $\sigma\to\sigma+x$ the Lagrangian reads
\begin{eqnarray}
  {\cal L}_{SU(2)} =&&-\frac{m^2_\varphi}2 x^2 -\frac{\lambda}4 x^2 -\sigma
  x(m^2_\varphi+\lambda x^2) + \bar \psi (i\dslash -m_\psi)\psi -
  g\bar\psi(\sigma+i\tau_i\pi_i)\psi 
  + \frac12 (\d_\mu\sigma)^2 -\frac{m_\sigma^2}2\sigma^2
  + \frac12 (\d_\mu\pi)^2 -\frac{m_\pi^2}2\pi^2
  \nn&&- \lambda x\sigma(\sigma^2+\pi_i^2)-
  \frac\lambda4(\sigma^2+\pi_i^2)^2,
\end{eqnarray}
where
\begin{equation}
  m_\psi^2=gx,\qquad m_\sigma^2=m_\varphi^2+3\lambda x^2,\qquad
  m_\pi^2=m_\varphi^2+\lambda x^2.
\label{Eq:SU2_masses}
\end{equation}

\subsection{The fermion wave function and $g$ renormalization}

We calculate the fermion self-energy on the background $x$:
\begin{eqnarray}
  \Sigma_\psi(k,x)= -\delta Z_\psi \slash\hspace*{-0.5em}k + \delta g
  x - ig^2 \pint4p \left[ iG_\sigma(p-k) i{\cal G}(p) -
    3iG_\pi(p-k) \gamma_5 i{\cal G}(p)\gamma_5 \right],
\end{eqnarray}
where $G_\sigma,$ $G_\pi,$ and ${\cal G}$ are the bosonic and fermion
propagators introduced in (\ref{Eq:propagators}) with the
corresponding masses given in (\ref{Eq:SU2_masses}). Taylor expanding
the bosonic propagators in $x$ and using that $\gamma_5$ anticommutes
with all the other gamma matrices, we find
\begin{equation}
  \Sigma_\psi(k) =\delta g x -\delta Z_\psi \slash\hspace*{-0.5em}k + 2
  ig^2 \pint4p \frac{2\slash\hspace*{-0.5em}p -
    m_\psi}{((p-k)^2-m_\varphi^2+i\ep)(p^2-m_\psi^2+i\ep)}.
\end{equation}
After evaluating the integrals, finiteness of the result requires the 
following expressions for the counterterms
\begin{equation}
  \delta Z_\psi = -\frac{g^2}{8\pi^2} \ln\frac{\Lambda^2}{\mu^2},\qquad
  \delta g = -\frac{g^3}{8\pi^2} \ln\frac{\Lambda^2}{\mu^2}.
\end{equation}

\subsection{The $\sigma$ mass, wave function and $\lambda$ renormalization}

Next, we calculate the $\sigma$ self-energy:
\begin{eqnarray}
  \Sigma_\sigma(k) =&& -\delta Z k^2 + \delta m_\varphi^2 +3\delta\lambda x^2+
  3\lambda \pint4p iG_\sigma(p) + 3\lambda \pint4p iG_\pi(p) - ig^2
  \pint4p\, \Tr\left[{\cal G}(p-k){\cal G}(p)\right]\nn&&
  +18i\lambda^2x^2 \pint4p G_\sigma(p-k)G_\sigma(p)+ 6i\lambda^2x^2
  \pint4p G_\pi(p-k)G_\pi(p).
\end{eqnarray}
The minus sign is because the fermionic bubble. After evaluating the
integrals we find, neglecting the $\Lambda^2$ corrections
\begin{equation}
  \Sigma_{\sigma}^{\mathrm{div}}(k) = -\delta Z k^2 +\delta m_\varphi^2+3\delta\lambda x^2 
  -\frac{3\lambda}{16\pi^2}\left(2m_\varphi^2+4\lambda x^2\right)
  \ln\frac{\Lambda^2}{\mu^2}+\frac{g^2}{4\pi^2}\left(\frac{k^2}6+3g^2x^2\right)
  \ln\frac{\Lambda^2}{\mu^2} - \frac{3\lambda^2}{2\pi^2}
  \ln\frac{\Lambda^2}{\mu^2}.
\end{equation}
Therefore, the expression of the counterterms is
\begin{equation}
  \delta Z = \frac{g^2}{24\pi^2} \,
  \ln\frac{\Lambda^2}{\mu^2},\qquad \delta m_\varphi^2 =
  \frac{3\lambda}{8\pi^2}\,m_\varphi^2 \ln\frac{\Lambda^2}{\mu^2},\qquad
  \delta\lambda = \frac{3\lambda^2-g^4}{4\pi^2}
  \ln\frac{\Lambda^2}{\mu^2}.
\end{equation}

\subsection{$\beta$-functions}

We again use the fact that the bare couplings are renormalization
group invariant:
\begin{equation}
  \frac{d g_0}{d\ln\mu} = \frac{d m_0^2}{d\ln\mu} =
  \frac{d\lambda_0}{d\ln\mu} =0,
\end{equation}
where
\begin{equation}
\label{Eq:chain_S}
  \frac d{d\ln\mu} =\frac\d{\d\ln\mu} + \beta_g \frac{\d}{\d g} +
  \beta_\lambda \frac{\d}{\d\lambda} + \gamma_\ph \frac{\d}{\d m_\ph^2}.
\end{equation}
This leads to:
\begin{equation}
  \beta_g = \frac{\d}{\d\ln\mu}\left[ \left(\delta
    Z+\frac12\delta Z_\psi\right)g - \delta g \right],\qquad \gamma =
  \frac{\d}{\d\ln\mu}\left[ \delta Z m_\varphi^2 - \delta  m_\varphi^2\right], \qquad
  \beta_\lambda = \frac{\d}{\d\ln\mu}\left[2\delta Z \lambda - \delta
    \lambda\right]. 
\end{equation}
Using the counterterms determined in the previous two subsections one has: 
\begin{equation}
  \frac{\d \delta Z_\psi}{\d\ln\mu} = \frac{g^2}{4\pi^2},\qquad
  \frac{\d \delta Z}{\d\ln\mu} = -\frac{g^2}{12\pi^2},\quad
  \frac{\d\delta g}{\d\ln\mu} =\frac{g^3}{4\pi^2},\quad
  \frac{\d\delta\lambda}{\d\ln\mu} = \frac{-3\lambda^2+g^4}{2\pi^2},\quad
  \frac{\d \delta m_\varphi^2}{\d\ln\mu} = - \frac{3\lambda m_\varphi^2}{4\pi^2}.
\end{equation}
Then, we find the following one-loop $\beta$-functions
\begin{equation}
  \frac{dg}{d\ln\mu}=\beta_g=-\frac{5g^3}{24\pi^2},\qquad
  \frac{d\lambda}{d\ln\mu} =\beta_\lambda =
  \frac{9\lambda^2-3g^4-\lambda g^2}{6\pi^2},\qquad
  \frac1{m_\ph^2}\frac{d m_\ph^2}{d\ln\mu} =\gamma =
  \frac{9\lambda-g^2}{12\pi^2}\,m_\varphi^2. 
\label{Eq:beta_fn_S}
\end{equation}

\section{Numerical strategy to solve the TCP equations}
\label{app:numstrat}

For numerical purposes it is advantageous to choose $\alpha$, $g^2=g^2(\mu)$ 
and $m_a$ as parametrization variables. Then we can proceed as follows. 
From \eqref{BT} we find
\begin{equation}
  u(\mu)=\frac{6\bar{\cal F}(\alpha)}{\pi^2}g^2,\qquad T^2 =
  \frac1{2g^2\left[\frac{g^2\bar{\cal F}(\alpha)}{\pi^2} +1
      +\frac{3}{\pi^2} \alpha^2\right]},
\end{equation}
then from \eqref{O4run}
\begin{equation}
  \Lambda_0^2 = \mu^2 e^{-\frac{24\pi^2}{5g^2(\mu)}} =
  \frac{e^{2\xi-\frac{24\pi^2}{5g^2(\mu)}}}
  {2g^2(\mu)\left[\frac{g^2(\mu)\bar{\cal F}(\alpha)}{\pi^2} +1 +
      \frac{3}{\pi^2} \alpha^2\right]}.
\end{equation}
Once we know $g^2(\mu),\,T,\,\Lambda_0$ and $u(\mu)$ we can compute
$m_\sigma$ by solving $m_\sigma^2=m_\sigma^2(\mu=m_\sigma)$ equation using
\eqref{mxrun}. Since now $m_\sigma(\mu)=1$ is the mass scale, 
in view of (\ref{mxrun}) we have to solve
\begin{equation}
  m_\sigma = \exp\left[\frac{7g_0^2}{24\pi^2}\!\!
    \int_0^{\ln\frac{m_\sigma}\mu} \!\! \frac{ds}{1+ b g^2(\mu)s}
     \, \frac{1+6 Y z(s)}{1-3 Y z(s)}\right],
\end{equation}
where $Y=(u(\mu)-1)/(3 u(\mu)+4).$
If we know $m_\sigma$ then we can have $g^2(m_\sigma)$ from
\eqref{O4run} which can be kept fixed.

From the running of $g$ \eqref{O4run} and $u$ \eqref{O4urun} we find
\begin{equation}
  g^2(m_a) = \frac{g^2}{1+\displaystyle\frac{5g^2}{12\pi^2}
    \ln\frac{m_a}\mu},\qquad
  \frac{2u(m_a)-1}{3u(m_a)+4}=\frac{2u(\mu)-1}{3u(\mu)+4}
  \left(1+\frac{\ln (m_a/\mu)}{\ln(\mu/\Lambda_{0})}\right)^{21/5}.
\end{equation}
From \eqref{scalerel} we find:
\begin{equation}
  \bar\Lambda_{0}^4 = 2 m_a^6 g^2(\mu)\left[\frac{g^2(\mu)\bar {\cal
        F}(\alpha)}{\pi^2} +1 + \frac{3}{\pi^2} \alpha^2\right]
  e^{\frac{24\pi^2}{5g^2(\mu)}}. 
\end{equation}
Having $\bar\Lambda_{0}^4,\, u(m_a)$ and $u(m_s)$ we can use the
solution of the $U(2)_L\times U(2)_R$ RG running \eqref{U2urun} to find
$m_s$. Finally from $m_s$ and $g(m_a)$ we compute from \eqref{U2run}:
\begin{equation}
  g^2(m_s)  = \frac{g^2(m_a)}{1-\displaystyle\frac{5g^2}{6\pi^2}
    \ln\frac{m_s}{m_a}}.
\end{equation}


\begin{thebibliography}{99}
\bibitem{deForcrand:2010ys}
  P.~de Forcrand,
  PoS {\bf LAT2009}, 010 (2009)
  [arXiv:1005.0539 [hep-lat]].

\bibitem{Fodor:2009ax}
  Z.~Fodor and S.~D.~Katz,
  arXiv:0908.3341 [hep-ph].

\bibitem{Fukushima:2010bq}
  K.~Fukushima and T.~Hatsuda,
  arXiv:1005.4814 [hep-ph].

\bibitem{Fodor:2004nz}
  Z.~Fodor and S.~D.~Katz,
  JHEP {\bf 0404}, 050 (2004)
  [arXiv:hep-lat/0402006].

\bibitem{Csikor:2004ik}
  F.~Csikor, G.~I.~Egri, Z.~Fodor, S.~D.~Katz, K.~K.~Szab{\'o} and A.~I.~Toth,
  JHEP {\bf 0405}, 046 (2004)
  [arXiv:hep-lat/0401016].

\bibitem{Ejiri:2003dc}
  S.~Ejiri, C.~R.~Allton, S.~J.~Hands, O.~Kaczmarek, F.~Karsch, E.~Laermann and C.~Schmidt,
  Prog.\ Theor.\ Phys.\ Suppl.\  {\bf 153}, 118 (2004)
  [arXiv:hep-lat/0312006].

\bibitem{Li:2010dy}
  A.~Li,
  ``Study of QCD critical point using the canonical ensemble method,''
  [arXiv:1002.4459 [hep-lat]].

\bibitem{Ejiri:2006ft}
  S.~Ejiri, T.~Hatsuda, N.~Ishii, Y.~Maezawa, N.~Ukita, S.~Aoki and K.~Kanaya,
  PoS {\bf LAT2006}, 132 (2006)
  [arXiv:hep-lat/0609075].

\bibitem{Gavai:2008zr}
  R.~V.~Gavai and S.~Gupta,
  Phys.\ Rev.\  D {\bf 78}, 114503 (2008)
  [arXiv:0806.2233 [hep-lat]].


\bibitem{Allton:2005gk}
  C.~R.~Allton {\it et al.},
  Phys.\ Rev.\  D {\bf 71}, 054508 (2005)
  [arXiv:hep-lat/0501030].


\bibitem{deForcrand:2008zi}
  P.~de Forcrand and O.~Philipsen,
  PoS {\bf LATTICE2008}, 208 (2008)
  [arXiv:0811.3858 [hep-lat]].

\bibitem{deForcrand:2006pv}
  P.~de Forcrand and O.~Philipsen,
  JHEP {\bf 0701}, 077 (2007)
  [arXiv:hep-lat/0607017].

\bibitem{deForcrand:2007rq}
  P.~de Forcrand, S.~Kim and O.~Philipsen,
  PoS {\bf LAT2007}, 178 (2007)
  [arXiv:0711.0262 [hep-lat]].


\bibitem{deForcrand:2010he}
  P.~de Forcrand and O.~Philipsen,
  arXiv:1004.3144 [hep-lat].

\bibitem{Casalbuoni:2006rs}
  R.~Casalbuoni,
  PoS C {\bf POD2006}, 001 (2006)
  [arXiv:hep-ph/0610179].

\bibitem{Berges:1998rc}
  J.~Berges and K.~Rajagopal,
  Nucl.\ Phys.\  B {\bf 538}, 215 (1999)
  [arXiv:hep-ph/9804233].

\bibitem{Scavenius:2000qd}
  O.~Scavenius, A.~Mocsy, I.~N.~Mishustin and D.~H.~Rischke,
  Phys.\ Rev.\  C {\bf 64}, 045202 (2001)
  [arXiv:nucl-th/0007030].

\bibitem{Jakovac:2003ar}
  A.~Jakov\'ac, A.~Patk{\'o}s, Zs.~Sz{\'e}p and P.~Sz{\'e}pfalusy,
  Phys.\ Lett.\  B {\bf 582}, 179 (2004)
  [arXiv:hep-ph/0312088].

\bibitem{Bowman:2008kc}
  E.~S.~Bowman and J.~I.~Kapusta,
  Phys.\ Rev.\  C {\bf 79}, 015202 (2009)
  [arXiv:0810.0042 [nucl-th]].


\bibitem{Kovacs:2006ym}
  P.~Kov\'acs and Zs.~Sz{\'e}p,
  Phys.\ Rev.\  D {\bf 75}, 025015 (2007)
  [arXiv:hep-ph/0611208].

\bibitem{Kovacs:2007sy}
  P.~Kov\'acs and Zs.~Sz{\'e}p,
  Phys.\ Rev.\  D {\bf 77}, 065016 (2008)
  [arXiv:0710.1563 [hep-ph]].

\bibitem{Schaefer:2008hk}
  B.~J.~Schaefer and M.~Wagner,
  Phys.\ Rev.\  D {\bf 79}, 014018 (2009)
  [arXiv:0808.1491 [hep-ph]].

\bibitem{Fukushima:2008wg}
  K.~Fukushima,
  Phys.\ Rev.\  D {\bf 77}, 114028 (2008)
  [Erratum-ibid.\  D {\bf 78}, 039902 (2008)]
  [arXiv:0803.3318 [hep-ph]].

\bibitem{Fukushima:2008is}
  K.~Fukushima,
  Phys.\ Rev.\  D {\bf 78}, 114019 (2008)
  [arXiv:0809.3080 [hep-ph]].

\bibitem{Chen:2009gv}
  J.~W.~Chen, K.~Fukushima, H.~Kohyama, K.~Ohnishi and U.~Raha,
  Phys.\ Rev.\  D {\bf 80}, 054012 (2009)
  [arXiv:0901.2407 [hep-ph]].

\bibitem{Gupta:2007dx}
  S.~Gupta,
  arXiv:0712.0434 [hep-ph].

\bibitem{Gupta:2008ac}
  S.~Gupta,
  J.\ Phys.\ G {\bf 35}, 104018 (2008)
  [arXiv:0806.2255 [nucl-th]].

\bibitem{Lenaghan:2000ey}
  J.~T.~Lenaghan, D.~H.~Rischke and J.~Schaffner-Bielich,
  Phys.\ Rev.\  D {\bf 62}, 085008 (2000)
  [arXiv:nucl-th/0004006].


\bibitem{Herpay:2006vc}
  T.~Herpay and Zs.~Sz{\'e}p,
  Phys.\ Rev.\  D {\bf 74}, 025008 (2006)
  [arXiv:hep-ph/0604086].


\bibitem{Philipsen:2005mj}
  O.~Philipsen,
  PoS {\bf LAT2005}, 016 (2006)
  [arXiv:hep-lat/0510077].


\bibitem{Collins} 
  J. Collins, 
  {\it Renormalization} (Cambridge Monographs
  for Mathematical Physics, Cambridge University Press, Cambridge, UK,
  1984)

\bibitem{'tHooft:1976fv}
  G.~'t Hooft,
  Phys.\ Rev.\  D {\bf 14} (1976) 3432
  [Erratum-ibid.\  D {\bf 18} (1978) 2199].

\bibitem{'tHooft:1976up}
  G.~'t Hooft,
  Phys.\ Rev.\ Lett.\  {\bf 37} (1976) 8.

\bibitem{PeskinSchroeder}
  M. E. Peskin and D. V. Schroeder,
  {\it An Introduction to Quantum Field Theory}, 
  (Westview Press, New York, 1995).

\end{thebibliography}
\end{document}